\documentclass[fleqn]{iopart}

\usepackage[all]{xypic}

\usepackage{vmargin}
\usepackage{ulem}
\usepackage{algorithm2e}
\usepackage{amsthm}
\usepackage{amsfonts,amssymb}
\usepackage[mathscr]{eucal}

\usepackage{algpascal}
\usepackage{color}
\usepackage{epsfig}

\usepackage{color}

\begin{document}

%% Title of the article. 
% The optional argument [] is the short version of the title,
% and the mandatory argument {} the title itself
\title[OT distances for seismic imaging]{A review of the use of optimal transport distances for high resolution seismic imaging based on the full waveform}

%% Authors, addresses and supports.
% The optional argument is for shortened version appearing in the headings. Please
% distinguish between first, middle and last names with the appropriate commands.

\author{{Ludovic} {M\'etivier}}
\address{Laboratoire Jean Kuntzmann\\
Universit\'e Grenoble Alpes\\
France}
% No support for the second author
%\thanks{}
%\email{ludovic.metivier@univ-grenoble-alpes.fr}
%
% Repeat the preceding commands for additional authors, commenting out lines
% which should not appear
\author{{Romain} {Brossier}}
\address{ISTerre\\
Universit\'e Grenoble Alpes\\
France}
% \thanks{The first author is supported by INRIA project OMEGA}
%\email{romain.brossier@univ-grenoble-alpes.fr}

\author{{F\'elix} {Kpadonou}}
\address{CGG\\
Massy\\
France}
%\email{felix.kpadonou@cgg.COM}
% Repeat the preceding commands for additional authors, commenting out lines
% which should not appear
\author{{J\'er\'emie} {Messud}}
\address{CGG\\
Massy\\
France}
%\email{jeremie.messud@cgg.com}

\author{{Arnaud} {Pladys}}
\address{ISTerre\\
Universit\'e Grenoble Alpes\\
France}

$\vspace{0cm}$

%% Keywords
\noindent{\it Keywords}: Optimal transport, convexity, optimization, seismic imaging
$\vspace{-0.5cm}$

\submitto{MathematicS In Action}

%% Mathematical classification   (2000)
%\subjclass{35R30,86A22,86A15}

%\graphicspath{
%% {/Data/GIT/ARTICLES/PUBLIS/2021_PLADYS_OT_VALHALL/},
%% {/Data/GIT/ARTICLES/PUBLIS/2016_METIVIER_TRANSPORT_IP/}
% {../2021_PLADYS_OT_VALHALL/},
% {../2016_METIVIER_TRANSPORT_IP/}
%}

% Abstract.
%\begin{abstract} 
%\end{abstract}

% Use the \maketitle command after the abstract
%\maketitle

%% Beginning of text

% Example of section
\section{Introduction}

This study is intended to review methodological developments done in the framework of high-resolution seismic imaging, based on a novel use of optimal transport distances. The high-resolution seismic imaging method considered here is called full-waveform inversion (FWI) in the geophysics community. 
FWI is a data fitting method aimed at inverting for subsurface mechanical parameters (mainly seismic wave velocities, but also density, attenuation, or anisotropy parameters).
Unlike tomography methods, which do not exploit the full data (or waveform) provided by the seismic recordings, but rather some extracted time-arrivals information, FWI aims to interpret the entire signal. The benefit is increased of resolution of the subsurface parameters reconstructed from the seismic data. While FWI was introduced in the early 1980s by French researchers in applied mathematics \cite{Lailly_1983_SIP} and geophysics \cite{Tarantola_1984_ISR}, its widespread adoption by the academic and industrial communities started in the past decade, supported by the development of wide-aperture/azimuth and broadband data acquisition schemes and parallel high-performance computing platforms. FWI is now applied at various scales: global, regional, and deep crustal scales in seismology, crustal and exploration scales in seismic imaging, and near surface scale in geotechnical engineering and archeology.\\ 

Despite this large adoption and many successful results, FWI still suffers from severe limitations. From a mathematical standpoint, FWI is a large scale PDE-constrained optimization problem. The misfit function that is used, which measures the discrepancy between observed seismic data and data calculated through the solution of a wave propagation problem, is non-convex. After discretization, the size of the FWI problem   (it is common to invert for millions of parameters) requires the use of local optimization solvers, which are prone to converge towards local minima.
This problem is all the more significant because of the nature of seismic data.
Thus, the success of FWI strongly depends on the choice of the initial model to ensure the convergence towards the global minimum of the misfit function.\\ 

This limitation, identified in the early days of FWI \cite{Gauthier_1986_TDN}, has been the motivation for a large variety of strategies. A short review of these strategies is proposed in Section 2.3. Among the different methods that have been investigated, the use of optimal transport (OT) distances-based misfit functions has been recently promoted \cite{Engquist_2014_WAS}. It has generated significant interest in the applied mathematics and geophysical communities, as the idea is elegant and the first application results were promising. The leading idea is to benefit from the inherent convexity of OT distances with respect to dilation and translation to render the FWI problem more convex.\\ 

However, the application of OT distances in the framework of FWI is not straightforward, as seismic data is signed, while OT has been developed for the comparison of probability measures.\\

The purpose of this study is to review two methods that were developed to overcome this difficulty. Both have been successfully applied to field data in an industrial framework. Both make it possible to better exploit the seismic data, alleviating the sensitivity to the initial model and to various conventional workflow steps, and reducing the uncertainty attached to the subsurface mechanical parameters inversion. In Section 2, we introduce the formalism of the FWI problem. We discuss its non-convexity and and provide a short review of conventional techniques designed to mitigate this non-convexity. In section 3, we detail our two propositions for the application of OT to seismic data. Numerical illustrations of these two methods on synthetic and field data are given in Section 4. Conclusion and perspectives finalize this study in Section 5. 

\newpage
\clearpage
\section{FWI: a non-convex PDE-constrained optimization problem}
\label{sec:FWI}

\subsection{Formalism and notations}

Here, we introduce the notations that will be used throughout the study. We start with the observed seismic data. Such data is generated by the recording of mechanical waves triggered by a seismic source. At global or regional scales, this source can be an earthquake occurring along a given fault. At smaller scales, which will be the main focus in this study, the source is controlled. Examples of controlled sources include an airgun in marine acquisition (offshore) or a vibrating truck in land acquisition (onshore). In a marine context, the receivers (or sensors) are deployed in the sea along cables towed by a boat (streamer acquisition) or at the sea bottom (node acquisition). For land data, the receivers are deployed at the Earth surface. Depending on the context, the receivers record the pressure variation (hydrophones) and/or the displacement in different directions (geophones, nodes). In the following, such observed data will be denoted by 
\begin{equation}
 d_{obs,s}(x_r,t) \in \mathcal{L}^{2}(\Sigma_r\times[0,T]), \;\; s=1,\dots,N_s.
\end{equation}
$\Sigma_r \subset \mathbb{R}^{d-1}$ denotes the Earth surface coordinates on which the receivers are deployed  (1 or 2-dimensional) and $T$ denotes the recording time (1-dimensional).
$d$ represents the total dimension of the representation (or data coordinate) space ($2$ or $3$).
$N_s$ denotes the number of seismic sources.\\ 

The calculated data, which are to be compared with the observed data, are obtained through the modeling of mechanical waves within the subsurface. Such waves can usually be modeled following the linear elasticity approximation, which considers the propagation of pressure waves (P-waves), shear waves (S-waves), and surface waves (Rayleigh and Love waves). In specific contexts, such as marine acquisition data, it is however possible to focus only on the propagation of P-waves under the acoustic approximation. In the following we introduce a general wave propagation operator $A(m)$ such that the wave equation we consider is denoted by
\begin{equation}
 A(m)u_s=b_s,
\end{equation}
where $m(x) \in \mathcal{L}^{2}(\Omega)$ represents the subsurface mechanical parameters with $\Omega \subset \mathbb{R}^{d}$, the dimensionality of the subsurface representation space being naturally considered to be the same as the dimensionality of the data representation space ($d=2$ or $3$).
$u_s(x,t) \in \mathcal{L}^{2}\left(\Omega \times [0,T]\right)$ is the wavefield solution of this wave equation and $b_s(x,t)\in \mathcal{L}^{2}\left(\Omega \times [0,T]\right)$ represents the seismic source term. In the following $m(x)$ will be referred to as the model parameter.\\   

The calculated data $d_{cal,s}[m](x_r,t) \in \mathcal{L}^{2}\left(\Sigma_r\times [0,T]\right)$ is defined for all $x_r\in \Sigma_r$ as 
\begin{equation}
 d_{cal,s}[m](x_r,t)=u_s[m](x_r,t),
\end{equation}
where the bracket $[m]$ is a reminder of the dependency of $d_{cal,s}$ and $u_s$ to the model parameter $m(x)$. In the following, we use a restriction operator $R$ to denote the relationship between $d_{cal,s}$ and $u_s$, such  that 
\begin{equation}
\begin{array}{cccc}
 R : &u_s &\longrightarrow& Ru_s=d_{cal,s}.
 \\
 &\mathcal{L}^{2}\left(\Omega \times [0,T]\right) & \longrightarrow& \mathcal{L}^{2}(\Sigma_r\times[0,T])
\end{array} 
\end{equation}
$R$ acts as a restriction of the wavefield space to the data space.

The general formulation for FWI is
\begin{equation}
\label{eq:fwi}
 \min_{m} f(m), 
\end{equation}
with 
\begin{equation}
 f(m)=\sum_{s=1}^{N_s}F (d_{cal,s}[m],d_{obs,s}),
\end{equation}
where $F(.,.)$ is a general positive function measuring the misfit between $d_{cal,s}$ and $d_{obs,s}$
\begin{equation}
\begin{array}{cccc}
 F : &(d_1,d_2) &\longrightarrow& F(d_1,d_2)
 \\
 &\mathcal{L}^{2}\left(\Sigma_r\times [0,T]\right)\times \mathcal{L}^{2}\left(\Sigma_r\times [0,T]\right) & \longrightarrow& \mathbb{R}_{+}
\end{array} 
\end{equation}

The conventional choice for $F$ is the least-squares misfit, such that
\begin{equation}
\label{eq:ls}
 F(d_1,d_2)=\frac{1}{2}\int_{\Sigma_r}\int_{0}^{T} |d_1(x_r,t)-d_1(x_r,t)|^2 dx_r dt
 ,
\end{equation}
leading to the difficulties mentioned in the introduction.

In this study, we discuss how OT distances can be advantageously introduced to define the operator $F$. Before discussing why the choice of a least-squares misfit yields a non convex function $f(m)$, we need to first take a detour to the numerical optimization strategy used to solve the problem \ref{eq:fwi}.\\
% and the computation of the gradient $\nabla f(m)$ using the adjoint state strategy.\\ 

As mentioned previously, the solution of \ref{eq:fwi} is performed using local optimization methods, which can be outlined as follows. Given an initial model $m_0$, such methods build a sequence
\begin{equation}
 m_{k+1}=m_k+\alpha_k \Delta m_k,
\end{equation}
where $\alpha_k \in \mathbb{R}^{+}_{*}$ is a scaling parameter computed by linesearch, and $\Delta m_k$ is a descent direction. In practice, we rely on quasi-Newton strategies, for which we have
\begin{equation}
 \Delta m_k=-Q_k\nabla f(m_k),
\end{equation}
where $\nabla f(m_k)$ is the gradient of the function $f(m)$ at $m_k$ and $Q_k$ is an approximation of the inverse Hessian of $f(m)$ at $m_k$ denoted by $H(m_k)^{-1}$
\begin{equation}
 Q_k\simeq H(m_k)^{-1}=\left(\nabla^2 f(m_k)\right)^{-1}.
\end{equation}
Usually, $Q_k$ is computed following the $l$-BFGS strategy (Brodyen-Fletcher-Goldfarb-Shanno formula), which builds a low-rank approximation of the inverse Hessian from gradients computed during the $l$-previous iterations \cite{Nocedal_1980_UQN}. More details on numerical optimization can be found in the reference book of Nocedal \cite{Nocedal_2006_NO}.\\

It is important to keep in mind that implementing a FWI algorithm requires the ability to compute $f(m)$ and its gradient $\nabla f(m)$. As a Jacobian-based computation of the gradient is computationally too expensive in practice (especially in terms of memory), the adjoint state strategy is usually employed \cite{Plessix_2006_RAS}. Following this method, the gradient of the total misfit function $f(m)$ is obtained as 
\begin{equation}
 \nabla f(m)= \sum_{s=1}^{N_s} \int_{0}^{T} \left(\frac{\partial A(m)}{\partial m}u_s[m]\right)(x,t)\lambda_s[m](x,t)dt,
\end{equation}
where $\lambda_s[m]$ is the wavefield solution of the adjoint equation
\begin{equation}
\label{eq:adj}
 A(m)^{T}\lambda_s=R^{T}\frac{\partial F}{d_{cal,s}}(d_{cal,s},d_{obs,s}).
\end{equation}
%and $^{T}$ denotes the adjoint.
This well-known result is derived in several studies; see for instance  \cite{Metivier_2016_OTI,Metivier_2019_GOT}.\\

Equation \ref{eq:adj} has a physical interpretation. The adjoint operator of the wave equation with an initial condition is the same wave equation with a final condition. Therefore the adjoint wavefield $\lambda_s$ is computed by a reverse propagation in time of the source term $R^{T}\frac{\partial F}{d_{cal,s}}(d_{cal,s},d_{obs,s})$. This source term is usually referred to as the adjoint source. Two contributions appear in the adjoint source: the first order derivative of the misfit function with respect to the calculated data and the adjoint of the restriction operator $R^{T}$. The latter operator acts as a lift from the data space to the wavefield space, yielding a source term localized at the receiver positions. The adjoint wavefield $\lambda_s$ is thus computed as the backpropagation of the adjoint source from the receiver positions. The final gradient is obtained as the summation over the sources of the zero lag correlation between the incident wavefield $u_s[m]$ (scaled by $\frac{\partial A(m)}{\partial m})$ and the adjoint field $\lambda_s[m]$.\\ 

Interestingly, we see from these formulas, and especially equation~(\ref{eq:adj}), that the only impact from a modification of the misfit measurement $F(.,.)$  is on the adjoint source definition. This is very convenient in terms of implementation, especially as the focus of this study is on introducing OT distances-based misfits. On the other hand, this also means that for each misfit function modification, one needs to be able to compute both the misfit measurement $F(.,.)$ and its first-order partial derivative $\frac{\partial F}{d_{cal,s}}(d_{cal,s},d_{obs,s})$. How to compute these quantities for OT distances-based misfits is an important question that will be discussed in Section \ref{sec:OT}.\\

Finally, one can note that for the least-squares misfit measurement, equation~\ref{eq:ls}, the adjoint source is simply
\begin{equation}
 \frac{\partial F}{d_{cal,s}}(d_{cal,s},d_{obs,s})= d_{cal,s}-d_{obs,s},
\end{equation}
which is the difference between calculated and observed data, also known as the residual. For a more developed physical interpretation of the gradient in FWI, the reader is redirected to   \cite{Operto_2013_TLE,Virieux_2017_FWI}.

% \begin{equation}
%  \forall d \in \mathcal{L}^{2}\left(\Sigma_r\times [0,T]\right), \;\; 
%  \|d\|^2=\int_{\Sigma_r}\int_{0}^{T} |d(x_r,t)|^2 dx_r dt.
% \end{equation}

% FWI is thus a nonlinear minimization problem where the least-squares misfit between observed and calculated data is minimized, over a space of model parameter $m(x)$. The relation between the calculated data and the model parameter are linked through a partial differential equation where $m(x)$ is a parameter while the calculated data is a restriction of its solution to a part of the surface where the receivers are deployed. 

\subsection{Non-convexity of least-squares based FWI}

The least-squares based FWI problem is notoriously non-convex \cite{Gauthier_1986_TDN}. The most widespread interpretation of this non-convexity is the following. The first-order parameters controlling the wave propagation within the subsurface are the seismic wave velocities. Perturbations of these parameters, provided that their spatial support is sufficiently large with respect to the wavelength of the propagated seismic waves, result mostly in time delays of the seismic waves. 
We are talking about sufficiently large-scale perturbations, or, equivalently, of sufficiently low-wavenumber perturbations in a Fourier domain interpretation.
In other words, the main difference between observed and calculated seismic wave packets is shifts in time, with a positive time-shift if the velocity decreases and a negative time-shift if the velocity increases. This effect has been carefully analyzed in the reference geophysics paper \cite{Jannane_1989_WES}. The point is that the least-squares misfit, which can be used to compare observed and calculated data, is not convex with respect to such time-shifts.\\ 

% A well known illustration of this non-convexity consists in the following Ricker wavelet test. A Ricker wavelet is a negative normalized second-order in time derivatives of a Gaussian (also known as Mexican hat wavelet), representative of a seismic signal in geophysics. In this test, a reference Ricker wavelet with a given central frequency is compared with a time-shifted Ricker wavelet with the same central frequency. The least-squares difference between the two is plotted against the time-shift. This plot exhibits two pathologic local minima and plateaus, and a narrow valley of attraction towards the global minimum. This non-convexity of the least-squares distance with respect to time-shifts is often referred to as cycle-skipping \cite{Virieux_2009_OFW}. It consists in erroneously putting in phase shifted wave packets, modifying the velocity in the opposite direction as what would be required to correctly align the phases of the seismic signal. This results in geological incorrect seismic velocity estimations.\\ 

This non-convexity with respect to time-shifts is often the main focus of FWI analysis
and is called the cycle-skipping issue in relation to the oscillatory nature of seismic data.
However, other seismic features may also affect the convexity of the problem: the sensitivity to the amplitude information present in the data (can affect the number of local minimums) or the quality of the low temporal frequency information present in the observed data (can affect the width of the global minimum valley).\\
%Interestingly, the choice of the misfit function can have an effect.

Finally, sources of non-uniqueness exist, which are related to the inability to predict the observed data with machine precision. This inability is due to seismic noise, which always contaminate the data, and the inaccuracy of the seismic wave modeling. The latter is partly due to uncertainty on the seismic source, which is difficult to estimate, in particular its coupling with the subsurface. Another source of uncertainty relies on the linear elasticity model used to simulate wave propagation itself, which is valid only in the small-displacement assumption. Attenuation effects are also difficult to predict and might play an important role depending on applications. Finally, the choice of the parameterization itself conditions the problem and the best one is often not trivial to determine. A compromise has to be found between having sufficient degrees of freedom to explain the data without introducing too much of a trade-off between the parameters that are reconstructed \cite{Operto_2013_TLE}. 

\subsection{Remedies to the non-convexity of least-squares based FWI}

To overcome the non-convexity issue, a standard remedy is to rely on a hierarchical workflow.
For the non-uniqueness issue, a remedy is to include regularization, which means introducing prior information into the problem to restrain the solution space.\\

A hierarchical workflow is a synonym for a multi-scale approach. The leading idea is to first interpret the low temporal frequency part of the data from a given initial model. The subsequent FWI result then serves as a new initial model for a new FWI step interpreting higher frequency data \cite{Bunks_1995_MSW}. This strategy is effective at reducing the cycle-skipping contribution to the non-convexity risk, as lower frequency data contains less propagated wavelength, thus reducing the risk of misaligning seismic travel-times. It can be complemented with time and offset windowing to focus the inversion on specific seismic events, further reducing the risk \cite{Shipp_2002_TDF,Wang_2009_RSW,Brossier_2009_SIC}.\\

Prior to the implementation of these hierarchical workflow, the initial model is designed with great care, usually through tomography methods that interpret arrival times of  specific seismic events \cite{Nolet_2008_BST}. The development of stereotomography methods has significantly improved the accuracy of these initial models \cite{Lambare_2008_STE}.\\ 

This conventional workflow (tomography + multiscale FWI) has been successfully applied to a large number of 2D and 3D datasets, at different scales, demonstrating the resolution power of FWI and its intrinsic interest for seismic imaging and subsurface characterization \cite{Fichtner_2010_FWT,Tape_2010_STS,Bozdag_2016_GAT,Plessix_2010_FWI,Stopin_2014_MWI,Operto_2015_ETF,Bretaudeau_2013_EFW,Groos_2014_RAF,Schafer_2013_TFW,Irnaka_2019_T3D}. 
However, situations exist which prevent the application of this workflow. The low-frequency part of the data might be too noisy to be interpreted. Picking arrival travel-times might be difficult because of noise or the presence of low-velocity anomalies in the shallow part of the Earth. This could cause the tomography-based initial model to be unreliable. In addition, even when the workflow can be applied with success, numerous steps, quality controls and human expertise are required, which in turn might question the robustness and the uncertainty attached to the estimated model. This has been the motivation for continuous effort to improve the robustness of FWI regarding the non-convexity issue, and to provide a better posed problem.\\ 

Reviewing such strategies goes beyond the scope of this study. Let us mention that they can roughly be divided into two categories: extension strategies and misfit function modifications.\\

The leading idea behind extension strategies is to introduce artificial degrees of freedom into the FWI problem to help fit the data in the early stage of the inversion. These degrees of freedom iteratively converge towards physical values during the FWI process. They can be introduced at the subsurface model level, following migration velocity analysis techniques \cite{Symes_2008_MVA}, or at the acquisition (source or receiver) level, as has been more recently proposed \cite{VanLeeuwen_2013_MLM,vanLeeuwen_2016_PMP,Aghamiry_2019_IWR,Huang_2019_WIS,Metivier_2021_REL}. From an optimization standpoint, introducing additional degrees of freedom can be seen as opening paths that connect local minima to the global minimum. These paths can be followed without moving uphill, thus using any local optimization solver.\\

On the other hand, misfit function modification appears as a more straightforward strategy. In the light of the non-convexity sources identified for the least-squares misfit function, it should be possible to improve the convexity of the FWI problem by modifying the way the misfit between observed and calculated data is computed. 
In particular, a misfit better than least-squares would be more sensitive to time-shifts and/or less sensitive to the amplitude information,
and/or would be better able to exploit the low-frequency information.
In other terms, it would put more weight on the kinematic information present in the data to 
mitigate the non-convexity.\\

Many propositions have been made in this direction, e.g., the use of instantaneous phase and envelope \cite{Fichtner_2008_TBC,Bozdag_2011_MFF,Wu_2014_SEI} or cross-correlation and deconvolution techniques \cite{Luo_1991_WET,VanLeeuwen_2010_CMC,Luo_2011_DBO,Warner_2016_AWI}.
Despite these attempts to design a better posed FWI problem, very few have been convincingly applied to 3D field data, with the exception of the acclaimed normalized deconvolution technique \cite{Warner_2016_AWI}. Most of these works remain at a conceptual level, with applications on sometimes too simplistic synthetic data examples.\\ 

More recently, the use of OT distances to compute the misfit between observed and calculated data has been promoted \cite{Engquist_2014_WAS}. The idea is to take advantage of the inherent convexity of optimal transport distances with respect to translation and dilation. In particular, designing a misfit function which should be convex with respect to time-shifts is a very appealing property, and a good proxy towards convexity with respect to wave velocities. Also, OT distances should make a global comparison of the seismic data possible, i.e., considering the data as a whole (beyond the pixelwise comparison induced by the use of the least-squares misfit), which could produce more convexity with respect to the amplitude information present in the data.\\

However, the path towards applications of OT distances to seismic data is not without difficulties. In particular, the OT theory has been developed in the frame of probability distributions, while seismic data consist in signed functions due to the oscillatory nature of the mechanical waves propagating in the subsurface, with varying ``mass'' (or integral of the data, which is especially true at low temporal frequencies).\\

To overcome this difficulty, different propositions have been made. The first consists of converting the seismic data to a probability distribution by a nonlinear transform and a normalization; for instance positive and negative part extractions \cite{Engquist_2014_WAS,Engquist_2016_WAS} or exponential/soft max encoding \cite{Yang_2017_AOT,Qiu_2017_EOT,Yang_2018_MRB,Yang_2018_WAS,Yang_2019_SEG}. While straightforward to apply, these methods present limitations for field data applications. The nonlinear transform is not easy to control, as it emphasizes specific parts of the data over others. This is detrimental to the inversion process and its stability. Sensitivity to noise and to the source function estimation can also be increased by such techniques.\\

To avoid these difficulties and apply OT distances to industrial field data, we have proposed two alternative strategies. The purpose of this study is to review these two techniques, and to illustrate their main features through applications to synthetic and field data.\\

The first of these two techniques relies on a specific dual form of the OT distance. This formulation has a close connection with the Kantorovich-Rubinstein norm \cite{Bogachev_2007_MET}, which is a well known tool in image processing. Its main benefits in the framework of FWI are its ability to consider the seismic data as a whole, taking into account the lateral coherency of the data in 2D or 3D representation spaces, to be less sensitive to the amplitude information, and to better exploit the low-frequency information in the data. These features enhance the convexity of the FWI problem.
However, as shown later, the enhancement of the convexity with respect to time-shifts specifically exists but remains limited when applied to signed data.\\

Improving even more the convexity with respect to time-shifts 
was the motivation for designing the second technique, named graph-space OT. In this framework, each 1D time-signal recorded by the receivers is considered, after time-discretization, as a point cloud in a 2D time/amplitude space. In terms of measure theory, this amounts to interpret each time-signal as a sum of 2D Dirac probability distributions. Standard OT distances can thus be computed and numerical tools dedicated to the comparison of point clouds through OT, arising from linear programming theory, can be advantageously employed. In doing so, we can greatly enhance the convexity with respect to time-shifts. 
However, each 1D time-signal is interpreted in a 2D time/amplitude space, increasing the computational cost, which for now excludes the possibility to consider the seismic data as a whole in 2D or 3D representation spaces.
Therefore, compared with the Kantorovich-Rubinstein norm strategy, the ability to exploit the lateral coherency of the data is lost,
along with the reduced sensitivity to the amplitude information and the enhancement of the low-frequency information in the data.
The two approaches thus complement each other, each working with different features that enhance the convexity of the problem.\\

In the next Section, after introducing notations and reminders about the OT theory, we present the formalism of the Kantorovich-Rubinstein norm and graph-space strategies.

\newpage
\clearpage
\section{Reformulating the full-waveform inversion problem using optimal transport distances}
\label{sec:OT}

\subsection{Generalities on optimal transport theory}

OT is a mathematical field originating from the work of the French mathematician Gaspard Monge \cite{Monge_1781_MSL} in 1781. The original problem was to minimize the efforts performed by workers to transfer sand piles to fill in holes on a bridge building site. The corresponding minimization problem formulated by Monge is not well posed, as a solution does not always exist. A well-posed relaxation of the problem was proposed by Kantorovich in 1942 \cite{Kantorovich_1942_TOM}. The solution of the OT problem, through the Kantorovich relaxation, defines a (Wasserstein) distance in the space of probability distributions.\\ 

Thanks to its convexity property with respect to translation and dilation, the OT distance has become widely used in image processing for applications such as image retrieval \cite{Rubner_2000_EMD,Rabin_2010_GSR}, histogram equalization \cite{Delon_2006_MVS}, color transfer \cite{Pitie_2007_ACG}, and texture mapping \cite{Dominitz_2010_TMO,Rabin_2012_WBA}. More references on image processing applications of OT can also be found in \cite{Lellmann_2014_KRU}.\\

In this section we recall the basic definition of the OT distance through the Kantorovich problem. We refer the readers to \cite{Villani_2008_OTO,Ambrosio_2003,Santambrogio_2015_OAM} for a more detailed introduction to the OT theory.\\

We start by recalling the standard Monge formulation. We consider two probability distributions $\mu \in \mathcal{P}(X)$ and $\nu \in \mathcal{P}(Y)$, where $X$ and $Y$ are measurable (here coordinate) spaces.
The push-forward distribution of $\mu\in\mathcal{P}(X)$ by the mapping $T$,
\begin{equation}
\left\{
\begin{array}{ccc}
\displaystyle
X&\longrightarrow& Y
\\
 T : x &\longrightarrow& T(x),
\end{array}
\right.
\end{equation}
is denoted by $T_{\#}\mu \in \mathcal{P}(Y)$, such that for any measurable set $A \subset Y$, we have
\begin{equation}
\label{eq:pushfwd}
 \left(T_{\#}\mu\right)(A)\equiv \mu\left(T^{-1}(A)\right)=\nu(A).
\end{equation}
Given a cost function $c(x,y)$ defined on data representation spaces, or ground cost,
\begin{equation}
\left\{
\begin{array}{ccc}
\displaystyle
X\times Y &\longrightarrow& \mathbb{R}_{+}
\\
\displaystyle
 c : (x,y) &\longrightarrow& c(x,y) ,
\end{array}
\right.
\end{equation}
the optimal transport problem is defined as 
\begin{equation}
\label{eq:MP}
 \min_{T} \; \left\{\int c(x,T(x)) d\mu(x), \;\;  T_{\#}\mu=\nu\right\}.
\end{equation}
(Note that the most general formulation is to use an ``$\inf$'' instead of the ``$\min$'' but both are equivalent with the real ground costs considered here.)
The constraint $T_{\#}\mu=\nu$ indicates that the push forward distribution $T_{\#}\mu$ of $\mu$ by the mapping $T$ is equal to the distribution $\nu$. The optimal transport problem can therefore be interpreted as determining the mapping $T$ that transports the distribution $\mu$ onto the distribution $\nu$ in the sense of equation (\ref{eq:pushfwd}), which minimizes the cost defined in (\ref{eq:MP}), for a given cost function $c(x,y)$. \\

The problem (\ref{eq:MP}) is difficult to solve, in particular because of the constraint (\ref{eq:pushfwd}). The Kantorovich relaxation of this problem takes the form of the following linear programming problem
\begin{equation}
\label{eq:KP}
 \min_{\gamma} \left\{\int_{X\times Y} c(x,y)d\gamma(x,y), \;\; u.c. \;\; \gamma \in \Pi(\mu,\nu) \right\},
\end{equation}
where the ensemble of transference plans $\Pi(\mu,\nu)$ is defined by 
\begin{equation}
\label{eq:KPcons}
 \Pi(\mu,\nu)=
\left\{ 
\gamma \in \mathcal{P}(X\times Y),\;\; \left(\pi_{X}\right)_{\#}\gamma=\mu, \;\: \left(\pi_{Y}\right)_{\#}\gamma=\nu
\right\}.
\end{equation}
The operators $\pi_{X}$ and $\pi_{Y}$ are the projectors on $X$ and $Y$, respectively. The problem (\ref{eq:KP}) generalizes (\ref{eq:MP}) in the sense that, instead of considering a mapping $T$ transporting each particle of the distribution $\mu$ to the distribution $\nu$, it considers all pairs $(x,y)$ of the space $X\times Y$ and for each pair defines how many particles of $\mu$ go from $x$ to $y$. In the context of the Monge formulation (\ref{eq:MP}), each point of the space $X$ has only one possible destination on $Y$, given by $T(x)$. In the context of the Kantorovich formulation (\ref{eq:KP}), the particles at point $x$ can have multiple destinations in $Y$, given by $\gamma(x,y)$ for $y\in Y$. The constraint (\ref{eq:KPcons}) ensures that the distribution $\mu$ is transported onto the distribution $\nu$. The relaxed problem (\ref{eq:KP}) admits a solution under very mild hypothesis, unlike Monge's problem (\ref{eq:MP}). In addition, when  (\ref{eq:MP}) admits a solution $T$, the measure $\gamma = ( I ,T)_{\#} \mu$ is a solution of the relaxed problem (\ref{eq:KP}) \cite{Ambrosio_2003,Pratelli_2007_MKP}.\\

The Kantorovich problem can be used to define a distance between probability measures, named $p$-Wasserstein distance, for $p\in \mathbb{N}^*$. We assume that 
\begin{equation}
 X=Y \subset \mathbb{R}^n, \;\; n \in \mathbb{N},
\end{equation}
and that the ground is induced by any norm $\|.\|$ put to power $p$, i.e.
\begin{equation}
 c(x,y)=\|x-y\|^p,
\end{equation}
and that $\mu$ and $\nu$ are probability measures with finite $p$-moment,
%$p\in \mathbb{N}$, such that
\begin{equation}
 \int_{X} \|x\|^pd\mu(x) < +\infty, \;\;
 \int_{X} \|x\|^pd\nu(x) < +\infty,
\end{equation}
with $\|.\|$ a norm on $\mathbb{R}^{n}$.
The $p$-Wasserstein distance between $\mu$ and $\nu$ is then defined as 
\begin{equation}
\label{eq:OT_problem}
W_p(\mu,\nu)=\left(\min_{\gamma \in \Pi(\mu,\nu)} \int_{X\times X} \|x-x'\|^p d\gamma(x,x')\right)^{1/p}.
\end{equation}

The convexity of the $p$-Wasserstein distance with respect to dilation and translation is a well-known result and has been analyzed in the context of seismic imaging in \cite{Engquist_2016_WAS}.\\

We see that the $p$-Wasserstein distance is defined for the comparison of probability measures. How to extend this problem in a mathematically consistent way to the comparison of signed measures is still an open question \cite{Ambrosio_2011_GFS,Mainini_2012_DTC}. 

\subsection{The Kantorovich-Rubinstein norm approach }

We have introduced the use of the Kantorovich-Rubinstein (KR) norm within FWI in two main studies. The first is dedicated to an audience of geophysicists, where the main concepts are introduced and several 2D synthetic applications are presented \cite{Metivier_2016_TOF}. The second is oriented towards an audience of applied mathematicians \cite{Metivier_2016_OTI}, where the formalism and connections with image processing are made, and the numerical strategy is further refined to obtain a linear/quasi-linear complexity solver to compute the KR norm in 2D and 3D, with a 3D synthetic application. Then, a wider audience paper for non-research geophysicists was published in The Leading Edge \cite{Metivier_2016_TLE}. 
Further publications involve a convexity analysis of the KR approach \cite{Metivier_2018_OTD} and an analysis of the KR norm FWI adjoint source properties \cite{Messud2021}.
We review here the main ideas outlined in these studies.

\subsubsection{Misfit function}

The KR norm approach is based on the $1$-Wasserstein distance. From equation~(\ref{eq:OT_problem}), we have
\begin{equation}
\label{eq:W1}
W_1(\mu,\nu)=\min_{\gamma \in \Pi(\mu,\nu)} \int_{X\times X} \|x-x'\| d\gamma(x,x').
\end{equation}

The following simplification of (\ref{eq:W1}) (using a dual formulation) can be obtained when $||.||$ is lower semi-continuous
\begin{equation}
\label{eq:W1dual}
\displaystyle
 W_1(\mu,\nu)=\max_{\varphi \in \textrm{Lip}_{1}} 
\int_{X} \varphi(x) d\left(\mu(x) -\nu(x)\right), \;\; 
\end{equation}
where $\textrm{Lip}_{1}$ denotes the space of $1$-Lipschitz function for the norm $\|.\|$, i.e.,
\begin{equation}
 \textrm{Lip}_{1}=
\left\{
\varphi: x \in X\longrightarrow \mathbb{R}, \;\; \forall (x,x')\; \in X\times X, \;\; \left|\varphi(x)-\varphi(x')\right| \leq \|x-x'\|
\right\}.
\end{equation}
The dual problem \ref{eq:W1dual} is a special instance of a more general duality result associated with the Kantorovich problem \ref{eq:KP} \cite{Santambrogio_2015_OAM}.\\

While the OT problem is defined for probability measures under its primal form (\ref{eq:W1}), the dual form \ref{eq:W1dual} can be extended for general measures $\mu$ and $\nu$ provided they have the same total mass (or integral), i.e., the mass is conserved from the mass distribution $\mu$ to the mass distribution $\nu$.\\ 

% An illustration of this requirement follows. Consider the case where $\mu$ and $\nu$ have a different mass
% \begin{equation}
%  \int_{X} d(\mu-\nu) \neq 0.
% \end{equation}
% Consider the constant function $\varphi_{\alpha}(x)=\alpha$ for $\alpha \in\mathbb{R}^{*}$. We have $\varphi_{\alpha} \in Lip_{1,c}$ and 
% \begin{equation}
%  \int_{X} \varphi_{\alpha} d\left(\mu -\nu\right) = \alpha \int_{X} d(\mu-\nu) \neq 0.
% \end{equation}
% Thus, the problem (\ref{eq:W1dual}) has no solution, as for any $\varphi \in Lip_{1}$, one can find $\varphi_{\alpha} \in Lip_{1}$ such that, for $\alpha$ sufficiently large, we have
% \begin{equation}
% \int_{X} \varphi_{\alpha} d\left(\mu -\nu\right) > \int_{X} \varphi d\left(\mu -\nu\right).
% \end{equation}

In addition, a straightforward generalization of the dual Kantorovich problem remains well posed even when the total mass between $\mu$ and $\nu$ is not the same. It complements the $1$-Lipschitz constraint with a bound constraint. This yields the distance
\begin{equation}
\label{eq:W1dual2}
\displaystyle
 W_{1,\lambda}(\mu,\nu)=\max_{\varphi \in \textrm{Lip}_{1}, \|\varphi\|_{\infty}< \lambda} 
\int_{X} \varphi(x) d\left(\mu(x) -\nu(x)\right).
\end{equation}

In the proposition made in \cite{Metivier_2016_TOF,Metivier_2016_OTI,Metivier_2016_TLE}, we focus on the particular case for which the norm $\|.\|$ on $X$ is actually the $\ell_1$ norm on $\mathbb{R}^{d}$ 
\begin{equation}
 \|x\|=
 %|x|=
 \sum_{i=1}^{n} |x_i|.
\end{equation}
Interestingly, with this choice, the generalization (\ref{eq:W1dual2}) corresponds to the definition of the KR norm \cite{Bogachev_2007_MET}. This norm is defined in the space of Radon measures on $X$, which is the dual space of the space of real valued continuous functions defined on $X$ that are zero at infinity for the $\|.\|_{\infty}$ norm, denoted by $\left(\mathcal{C}_0\left(\Omega,\mathbb{R}),\|.\|_{\infty}\right)\right)$. Besides the link with OT, the KR norm can also be interpreted as a generalization of the $L^1$ norm (in a similar sense of the generalization from Total Variation to Total Generalized Variation norms) and shares some properties with the Meyer's G-norm. These similarities are studied in detail in \cite{Lellmann_2014_KRU}, where the use of the KR norm is proposed as an alternative to the $L^1$ norm in a Total Variation denoising problem. 
%In the remainder of the paper, the space of $1$-Lipschitz functions for the distance induced by the $\ell_1$ norm on $\mathbb{R}^{d}$ and with infinity norm bounded by $\lambda$ will be denoted by $\textrm{BLip}_{1,\lambda}$.\\

More generally, a Mahalanobis-like $\ell_1$ norm, 
\begin{equation}
\label{eq:l1_weights}
 \|x\|=
 %|x|=
 \sum_{i=1}^{n} \frac{1}{\sigma_i}|x_i|,
\end{equation}
must be used as soon as the dimensionality of the various axes of the space $X$ do not have the same physical dimensions.
The $\sigma_i$, with $\infty>1/\sigma_i>0$, then denote standard-deviation-like weights
that can rescale the different physical dimensions and account for uncertainties.
This is important in seismic, where $n=d$ (=2 or 3) with $X=\Sigma_r\times[0,T]$ (1 or 2 distance coordinates and 1 time coordinate), following the notations introduced in Section \ref{sec:FWI}.
The benefit of adding these weights has been studied in \cite{Messud2019,Messud2021}.

\subsubsection{Adjoint source}

In the frame of seismic where $X=\Sigma_r\times[0,T]$, using the notations introduced in Section~\ref{sec:FWI}, we propose the following KR norm-based misfit for FWI 
%$X=\Sigma_r\times[0,T]$ and
\begin{equation}
 F(d_{cal,s},d_{obs,s})= W_{1,\lambda}(d_{cal,s},d_{obs,s})=
 \max_{\varphi \in \textrm{Lip}_{1}, \;\; \|\varphi\|_{\infty}< \lambda} 
\int_{\Sigma_r} \int_{0}^{T}\varphi(x_r,t) \left(d_{cal,s}(x_r,t) -d_{obs,s}(x_r,t)\right) dx_r dt.
\end{equation}
As mentioned previously, we need to access the quantity
 \begin{equation}
 \frac{\partial F}{\partial d_{cal,s}}(d_{cal,s},d_{obs,s})= \frac{\partial W_{1,\lambda}} {\partial d_{cal,s}}(d_{cal,s},d_{obs,s}).
\end{equation}
We denote the solution of (\ref{eq:W1dual2}) by $\overline{\varphi}$, such that 
 \begin{equation}
  \overline{\varphi} = \arg\max_{\varphi \in \textrm{Lip}_{1}, \;\; \|\varphi\|_{\infty}< \lambda} 
\int_{\Sigma_r} \int_{0}^{T}\varphi(x_r,t) \left(d_{cal,s}(x_r,t) -d_{obs,s}(x_r,t)\right) dx_r dt.
 \end{equation}
Using the almost-everywhere (a.e.) differentiability of concave functions, we have for a.e. $d_{cal}$
\begin{equation}
\label{eq:resKR}
\frac{\partial F} {\partial d_{cal,s}}(d_{cal,s},d_{obs,s})=\overline{\varphi} \;\; \textrm{a.e.}
\end{equation}
This result shows that the implementation of the KR approach in the framework of FWI requires a single numerical method to solve the problem \ref{eq:W1dual2}. The maximum value of the criterion in the definition of \ref{eq:W1dual2} provides the misfit function value $F(d_{cal,s},d_{obs,s})$. The function $\overline{\varphi}$ reaching this maximum provides the adjoint source required for the gradient computation.\\

$\overline{\varphi}$ can be conceptualized as the result of a smart processing of the least-squares adjoint source $d_{cal,s} -d_{obs,s}$. The Lipschitz and bound constraints will tend to reduce the dynamics of amplitudes and enhance the low frequency content present in $d_{cal,s} -d_{obs,s}$
(together with producing wavelets that will tend to become piecewise linear a.e. it matters for the FWI problem). 
Also, the KR adjoint-source can enhance the lateral continuity of events in $d_{cal,s} -d_{obs,s}$ using properly tuned weights in (\ref{eq:l1_weights}).
These elements are formally studied in \cite{Messud2021}.

\subsubsection{Numerical computation}
To keep the presentation compact, we give the technical details of the algorithm we set up for the numerical solution of the problem \ref{eq:W1dual2} in Appendix \ref{app:KR}. Let us mention here that the algorithm inherits a linear or quasi-linear complexity from the combination of three elements: a reduction of the number of constraints from the use of a $\ell_1$ norm as a ground cost, exploiting the ``Manhattan'' property of the $\ell_1$ norm; the use of the proximal splitting algorithm ``Alternative direction method of multipliers'' (ADMM) solver; the identification of the linear system to solve at each ADMM iteration as a finite-difference discretized Poisson's problem for which efficient solvers exists (either based on Fast Fourier transform or multigrid strategies). \\

In seismic, (\ref{eq:W1dual2}) can be resolved considering different effective dimensionalities for the data representation space $X$ (using previously introduced notations).
The 3D case, i.e., $X=\Sigma_r\times[0,T]$ with $\Sigma_r \subset \mathbb{R}^{2}$, tends to be too costly for industrial applications because of the size of the linear problem that is expressed in Appendix \ref{app:KR}.
As a consequence, most 3D data applications first split the data into 2D receiver ``lines'' considering
$\Sigma_r=\Sigma_r^{line}\times\Sigma_r^{\perp}$ with $\Sigma_r^{line}\subset \mathbb{R}$ and $\Sigma_r^{\perp}\subset \mathbb{R}$; then, the KR problem is solved independently for each line, i.e. considering $X\rightarrow\Sigma_r^{line}\times[0,T]$ in (\ref{eq:W1dual2}). With an effective dimensionality reduced to 2, the KR problem becomes manageable in an industrial context.
In practice, the direction of the receiver lines is chosen to be the best sampled (or less noisy) one.

\subsection{The graph-space approach}

The graph-space OT concept was first introduced in a paper published in the GEOPHYSICS journal \cite{Metivier_2018_OTM}. However, in this preliminary study, the underlying computational cost was too expensive for possible applications in realistic settings. Only the analysis and the development of the associated numerical strategy, performed in \cite{Metivier_2019_GOT} and recently refined in \cite{Metivier_2021_NGS}, have made such applications possible. We review here the main ideas presented in these studies. 

\subsubsection{Misfit function}

Whereas the KR norm implementation considers the seismic data as a ``whole'', i.e., $X\rightarrow\Sigma_r^{line}\times[0,T]$ in computations done independently for each defined line in $\Sigma_r$,
the graph-space implementation considers the data as a collection of 1D time functions or ``traces'', i.e., $X\rightarrow[0,T]$ in computations done independently for each position in $\Sigma_r$.
Each 1D time function is denoted by
\begin{equation}
 d_{cal,s,r}(t), d_{obs,s,r}(t), \;\; s=1,\dots,N_s, \;\; r=1,\dots,N_r,
\end{equation}
with
\begin{equation}
d_{cal,s,r}(t)=d_{cal,s}(x_r,t),\;\;
d_{obs,s,r}(t)=d_{obs,s}(x_r,t).   
\end{equation}
This amounts to a discretization of the receiver variable $x_r$, considered as continuous in the previous sections.\\

For the sake of clarity, let us consider for a moment the simple case where $N_s=1$ and $N_r=1$, and drop the subscript $s$ and $r$. After subsequent time discretization, each function $d_{cal}(t)$ and $d_{obs}(t)$ can be considered as discrete point clouds in a 2D time/amplitude space, or graph-space.
Assuming that the time discretization is the same for both $d_{cal}(t)$ and $d_{obs}(t)$, which is satisfied in practice, %and assimilating $d_{cal}$ and $d_{obs}$ with their discrete point clouds with a slight abuse of notations, 
we consider
\begin{equation}
\begin{array}{l}
 \displaystyle
d_{cal}^{GS} \in \left(\mathbb{R}^2\right)^{N_t}, \;\; d_{cal}^{GS}=\{\left(t_i,d_{cal}(t_i)\right), \;\; i=1,\dots,N_t\},\;\;
\\
\displaystyle
d_{obs}^{GS} \in \left(\mathbb{R}^2\right)^{N_t}, \;\; d_{obs}^{GS}=\{\left(t_i,d_{obs}(t_i)\right), \;\; i=1,\dots,N_t\}, 
\end{array}
\end{equation}
where $N_t \in \mathbb{N}$ is the number of time samples.
We now associate a Dirac probability density function to each point of the discrete point clouds $d_{cal}^{GS}$ and $d_{obs}^{GS}$ and compute the corresponding $p$-Wasserstein distance using (\ref{eq:OT_problem}). A standard result shows that computing such a $p$-Wasserstein distance to power $p$ is equivalent to solving the following linear sum assignment problem:
\begin{equation}
 W_{p}^p(d_{cal}^{GS},d_{obs}^{GS})=\min_{\sigma \in S(N_t)}\sum_{i=1}^{N_t} 
 c_{i,\sigma(i)}\left(d_{cal}^{GS},d_{obs}^{GS}\right)
 ,
\end{equation}
where $S(N_t)$ is the ensemble of permutation of $\left\{1,\dots,N_t\right\}$ and
the cost in the graph representation space can be defined by
%$c_{ij}\left(d_{cal},d_{obs}\right)$ is defined by 
\begin{equation}
 c_{ij}\left(d_{cal}^{GS},d_{obs}^{GS}\right)
 =||\left(t_i,d_{cal}(t_i)\right)-\left(t_{j},d_{obs}(t_j)\right)||_p^p
 =|t_i-t_j|^p + |d_{cal}(t_i)-d_{obs}(t_j)|^p.
\end{equation}
A proof of this equivalence can be found in \cite{Villani_2003_AMS}.\\

In the following, we consider a weighted version $c_{ij}^{\eta}$ of $c_{ij}$ such that 
\begin{equation}
 \label{eq:cij}
 c_{ij}^{\eta}\left(d_{cal}^{GS},d_{obs}^{GS}\right)=\eta|t_i-t_j|^p + |d_{cal}(t_i)-d_{obs}(t_j)|^p,
\end{equation}
where $\eta \in \mathbb{R}^{+}$ is a dimensioning parameter whose role is discussed later.
The defined $p$-Wasserstein distance power $p$ is minimized when the point clouds $d_{cal}^{GS}$ and $d_{obs}^{GS}$ coincide,
thus when the functions $d_{cal}(t_i)$ and $d_{obs}(t_i)$ coincide, which is in agreement with our goal.
As the time sampling and values $t_i$ are fixed in our applications,
we can consider the $p$-Wasserstein distance power $p$ to be a function of $d_{cal}(t_i)$ and $d_{obs}(t_i)$ only, and finally have
\begin{equation}
\label{eq:gGSOT}
 g\equiv W_{p}^p(d_{cal},d_{obs})=\min_{\sigma \in S(N_t)}\sum_{i=1}^{N_t} 
 \left(\eta|t_i-t_{\sigma(i)}|^p + |d_{cal}(t_i)-d_{obs}(t_{\sigma(i)})|^p\right).
\end{equation}
%Amplitude displacement along the time axis is then accounted for by the minimizing permutation.

Re-introducing the source and receiver dependencies, the misfit function $F(d_{cal,s},d_{obs,s})$ is obtained by summing the various receiver contributions
\begin{equation}
\label{eq:FGSOT}
 F(d_{cal,s},d_{obs,s})=\sum_{r=1}^{N_r} g(d_{cal,s,r},d_{obs,s,r})
\end{equation}
% where 
% \begin{equation}
% \label{eq:gGSOT}
%  g(d_{cal,s,r},d_{obs,s,r})=\min_{\sigma \in S(N_t)}\sum_{i=1}^{N_t} \left(\eta|t_i-t_{\sigma(i)}|^p + |d_{cal,s,r}(t_i)-d_{obs,s,r}(t_{\sigma(i)})|^p\right).
% \end{equation}

\subsubsection{Adjoint source}

We have from \ref{eq:FGSOT}
\begin{equation}
 \label{eq:gGSOT_adjs}
 \frac{\partial F}{\partial d_{cal,s,r}}(d_{cal,s},d_{obs,s})=
 %\sum_{r=1}^{N_r}
 \frac{\partial g}{\partial d_{cal,s,r}}(d_{cal,s,r},d_{obs,s,r}).
\end{equation}
This calls for a definition of the quantity $\frac{\partial g}{\partial d_{cal,s,r}}(d_{cal,s,r},d_{obs,s,r})$. In \cite{Metivier_2019_GOT}, we prove the following result
\begin{equation}
 \frac{\partial g}{\partial d_{cal,s,r}}(d_{cal,s,r},d_{obs,s,r})=
 p|d_{cal,s,r}-d_{obs,s,r}^{\overline{\sigma}}|^{p-2}\left(d_{cal,s,r}-d_{obs,s,r}^{\overline{\sigma}}\right)\;\; a.e.
\end{equation}
where $\overline{\sigma}$ is defined by
\begin{equation}
 \overline{\sigma}=\arg\min_{\sigma \in S(N_t)}\sum_{i=1}^{N_t} \left(\eta|t_i-t_{\sigma(i)}|^p + |d_{cal}(t_i)-d_{obs}(t_{\sigma(i)})|^p\right),
\end{equation}
and 
\begin{equation}
 d_{obs,s,r}^{\sigma}(t_i)=d_{obs,s,r}\left(t_{\sigma(i)}\right), \;\; i=1,\dots,N_t.
\end{equation}
This result calls for several comments. First, within the graph-space approach, the $p$-Wasserstein distance power $p$ between observed and calculated data can be seen as a generalization of $L^p$ distances power $p$. Indeed, the adjoint source associated with the latter would be given by
\begin{equation}
 \frac{\partial g}{\partial d_{cal,s,r}}(d_{cal,s,r},d_{obs,s,r})= p|d_{cal,s,r}-d_{obs,s,r}|^{p-2}\left(d_{cal,s,r}-d_{obs,s,r}\right),
\end{equation}
with $p=2$ being the least-squares case.
The generalization to the graph-space OT adjoint-source (\ref{eq:gGSOT_adjs}) comes from the optimal assignment $\overline{\sigma}$ computed as the solution of the OT problem between the seismic data traces seen as point clouds. Instead of comparing calculated and observed traces at the same time samples $t_i, \;\; i=1\dots,N_t$, they are compared at time samples matched by this optimal assignment, which are $t_i$ and $t_{\overline{\sigma}(i)}, \;\; i=1\dots,N_t$.\\

Second, as in the KR approach, implementing the graph-space strategy within FWI requires a single solver, computing the solution of \ref{eq:gGSOT}. For a single trace $s,r$, the minimum value of the criterion in \ref{eq:gGSOT} provides the misfit function value, while the adjoint source can be determined from the optimal assignment $\overline{\sigma}$ achieving this minimum value. The final misfit function $F\left(d_{cal,s},d_{obs,s}\right)$ and its adjoint source 
are obtained considering all the traces.

\subsubsection{Choice of the parameter $\eta$}

In \ref{eq:cij}, the scaling parameter $\eta$ plays a crucial role. It controls the behavior of the permutation $\overline{\sigma}$ (and thus of the graph-space misfit function $g(d_{cal,s,r},d_{obs,s,r})$) by weighting the cost of assigning points of the graphs of $d_{cal,s,r}$ and $d_{obs,s,r}$ along the time axis. If $\eta$ is chosen to be ``large'', the assignment is preferably done along the amplitude axis, and the graph-space misfit function boils down to the conventional least-squares misfit. On the other hand, if $\eta$ is ``small'', the assignment is preferably done along the time axis, and the graph-space misfit function becomes sensitive to time shifts.\\ 
 
More precisely, a practical choice for $\eta$ is, for a trace $s,r$,
 \begin{equation}
\label{eq:eta}
 \eta \equiv  \eta_{s,r}=\frac{A_{s,r}^2}{\tau^2},
\end{equation}
where $\tau \in \mathbb{R}$ is a maximum expected time shift and $A_{s,r} \in \mathbb{R}$ is an amplitude normalization parameter, for instance the difference between the maximum amplitude peaks in $d_{cal,s,r}$ and $d_{obs,s,r}$. Following this definition, a point $\left(t,d_{cal,s,r}(t)\right) \in \mathbb{R}^2$ such that $d_{cal,s,r}(t)-d_{obs,s,r}(t) \in \mathbb{R}$ is equal to $A_{s,r}$ can be assigned with a point of the same amplitude but shifted in the time-direction by $\tau'$ such that $|\tau'|\leq\tau$. If $|\tau'|>\tau$ then it will be assigned with $\left(t,d_{obs,s,r}(t)\right)$. As is illustrated in the next Section, this scaling strategy provides a convenient way to calibrate the graph-space misfit function, to make it convex with respect to time-shifts as large as $\tau$.  

\subsubsection{Numerical computation}

Numerous economy field problems can be modeled as linear sum assignment problems. For this reason, various algorithms have been proposed during the second half of the twentieth century, see for instance \cite{Bertsekas_1998_NOC,Burkard_2012_APR} for a review. These algorithms can be divided in three main classes: those based on primal-dual methods (among them the Hungarian algorithm \cite{Kunh_1955_HMA}); those based on a specification of the simplex algorithm, either the primal \cite{Akgul_1993_GPP} or dual \cite{Balinski_1985_SMA} version of the simplex method; those based on purely dual algorithms, a category to which belongs the auction strategy introduced by \cite{Bertsekas_1989_AAT}. From different studies \cite{Bertsekas_1998_NOC,Burkard_2012_APR}, it appears that the auction algorithm, combined with an $\varepsilon$-scaling technique, achieves one of the best worst-case complexity. Benchmarking experiments on different sets of reference problems also show its good performance for the solution of small scale dense problems.\\

In our applications in the frame of seismic imaging, the observed complexity of the auction algorithm is between quadratic and cubic, and the computation time to solve instances of problems with point clouds containing up to one thousand points is very small (typically less than 1 second on a single core architecture). 
This is within the order of the number of time samples one has to consider to represent a single seismic trace at Nyquist sampling.
%Considering total time duration and frequency content of the seismic data, at Nyquist sampling. 
For this reason, the auction algorithm has proven very useful for our applications. A full description of the algorithm is beyond the scope of this study. We refer the interested reader to  \cite{Bertsekas_1989_AAT,Bertsekas_1998_NOC,Burkard_2012_APR,Metivier_2019_GOT} for a complete presentation of the auction algorithm.

\newpage 
\clearpage
\section{Illustration on synthetic and field data examples}

In this Section, we illustrate the main properties of the KR and the graph-space approaches in the framework of FWI and present applications of these two strategies to 3D field data. Let us mention that, from a methodological point of view, these two methods have been compared to each other, and also with more conventional strategies to mitigate non-convexity in FWI mentioned in Section 2.3. This comparison has been the main topic and motivation of a recently published paper in the journal GEOPHYSICS, which might be of interest for the reader \cite{Pladys_2021_OCM}.

%Let us mention that, from a methodological point of view, these two methods have been compared to each other, and also to the use of other misfits mentioned in the introduction, in a much more extensive way, which might be of interest for the reader \cite{Pladys_2021_OCM}.

\subsection{A simple Ricker synthetic test to illustrate the convexity with respect to a time-shift}

\begin{figure}[!ht]
 \begin{center}
  \includegraphics[scale=0.5]{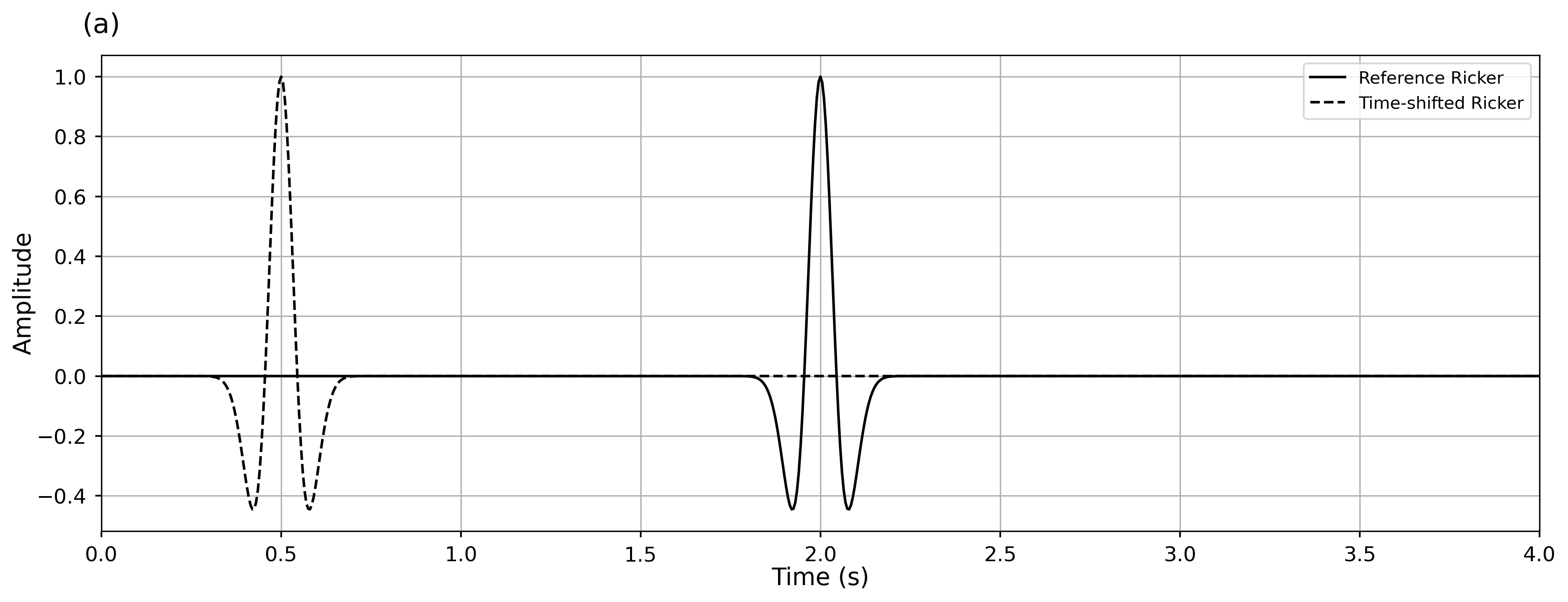}
  \includegraphics[scale=0.5]{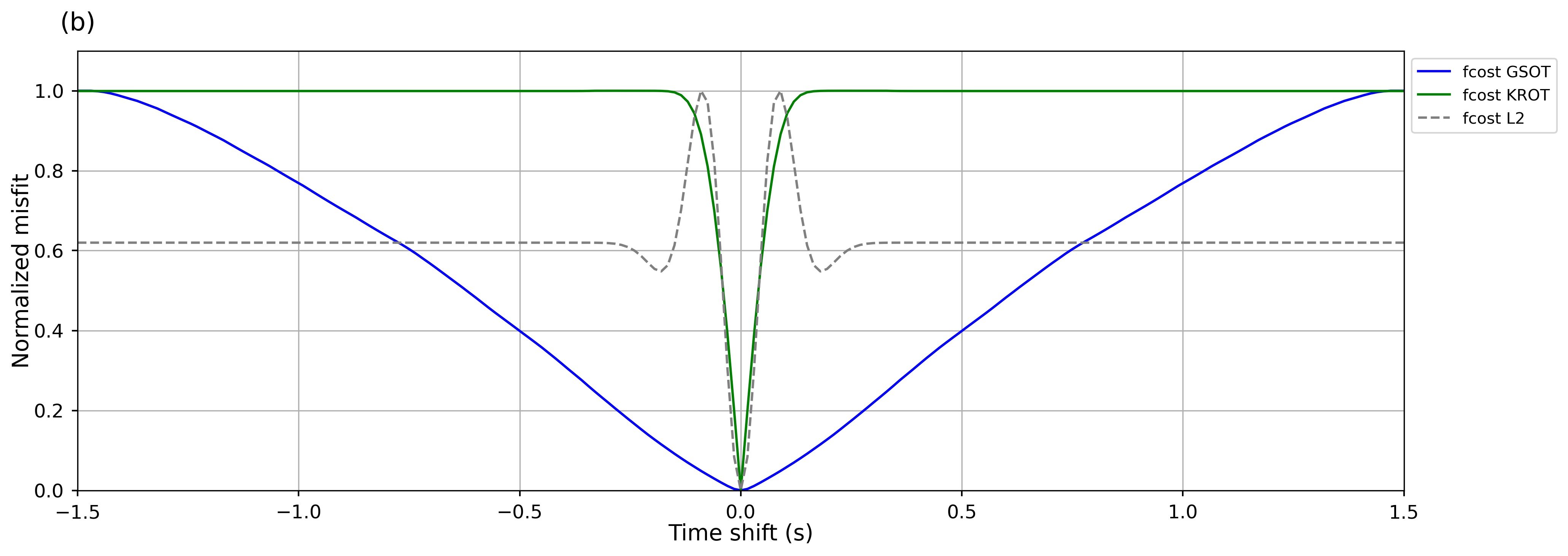}
  \caption{Comparison of the least-squares, KR and graph-space misfit values for 2 shifted Ricker wavelets.}
  \label{fig:Ricker}
 \end{center}
\end{figure}

We firstly illustrate the convexity properties of the two approaches with respect to a time-shift, related to the cycle-skipping issue. Ricker-type time wavelets are considered here. Such wavelets, also known as Mexican hat wavelets, are commonly used in geophysics to represent seismic sources. Mathematically, a Ricker corresponds to a second-order derivative of a Gaussian, and can be expressed as 
\begin{equation}
 r[t_0,f_0](t)=\left(1-2\pi^2 f_0^2(t-t_0)^2\right)\exp\left(-\pi^2 f_0^2(t-t_0)^2\right).
 \label{eq:ricker}
\end{equation}
In \ref{eq:ricker}, $f_0$ is the central frequency in Hertz and $t_0$ is a time delay in seconds such that the Ricker wavelet peak (or maximum) is at $t_0$. We consider a reference Ricker wavelet $r_{ref}(t)$, such that
\begin{equation}
 r_{ref}(t)=r[2,5](t),
\end{equation}
on a time interval $[0,T]$ with $T=4$ s. We then build a series of Ricker wavelets shifted in time $r_{shift}(t)$ such that
\begin{equation}
 r_{shift}[s](t)=r[2+s,5](t), \;\; s \in [-1.5, 1.5].
\end{equation}
The shifted Ricker wavelets have the same shape as the reference wavelet (same central frequency of 5 Hz). The Ricker wavelets $r_{ref}(t)$ and $r_{shift}[-1.5](t)$ are presented in Figure \ref{fig:Ricker}a. Then, for each time shift $s$, we compute the distance between $r_{ref}$ and $r_{shift}[s]$ using the KR approach and the graph-space OT approach. Namely, according to previously introduced notations, we compute the following functions of $s$:
\begin{equation}
 W_{1,\lambda}\left(r_{shift}[s],r_{ref}\right),\;\;  
 g\left(r_{shift}[s],r_{ref}\right),\;\;  s \in [-1.5, 1.5].
\end{equation}
$\lambda$ is chosen equal to $1$, while for the graph-space OT approach the $\tau$ parameter is set to $\tau=1.5$ s. The results are presented in Figure  \ref{fig:Ricker}b together with what would be obtained following a standard least-squares approach.\\

As can be seen in Figure \ref{fig:Ricker}, the least-squares approach produces a multi-modal misfit function, with a global minimum reached for $s=0$ (no time-shift), and two local minima reached approximately at $s=-1.15$ s and $s=1.15$ s. The left local minimum corresponds to the situation in which the right side-lobe of $r_{shift}$ is in phase with the left side-lobe of $r_{ref}$. The right local minimum corresponds to the opposite situation. When the two Ricker wavelets do not overlap, the least-squares misfit becomes constant. This is an illustration of the non-convexity of the least-squares misfit function with respect to a time-shift. In an FWI analogy, for the method to converge towards a meaningful subsurface model, one would need an initial model predicting the data within a time-shift between approximately $-0.1$ and $0.1$ s; otherwise, the method would converge towards a local minimum or would stagnate at the initial estimation.\\

The KR and graph-space strategies exhibit different misfit function profiles. Both present a single global minimum. The KR approach improves to some extent the pathologies associated with the convexity of the least-squares misfit, the KR misfit function exhibiting a wider valley of attraction. This brings more robustness to cycle-skipping, especially when dealing with the low temporal frequencies of the data \cite{Messud2021}. In an FWI analogy, to make the KR method converge towards a meaningful subsurface model, one would need an initial model predicting the data within a time shift between $-0.15$ and $0.15$s. However, the KR misfit function exhibits two large regions where the misfit function is almost constant. The reason for this loss of convexity with respect to time-shifts has been documented for instance in \cite{Metivier_2018_OTD}. It can be shown that computing $W_{1,\lambda}$ for signed data is equivalent to summing the OT distance between the negative (respectively, the positive) part of the calculated data and the positive (respectively, the negative) part of the observed data. This decomposition has been proposed by Mainini \cite{Mainini_2012_DTC} to  extend OT distances to signed data. However, such decomposition does not provide a measure of distance that is convex with respect to time shifts.\\

Conversely, the graph-space strategy produces a misfit function that is monotonically decreasing and close to being convex with respect to the time-shift, that is the expected behavior with the choice of $\tau=1.5$ s (which is the maximum absolute time shift considered here).\\

This simple example illustrates the pathological behavior of the least-squares approach,
the interest of using OT-based misfits to enhance the convexity with respect to time-shifts,
and the superiority of the graph-space approach over the KR approach regarding this convexity.
We now illustrate that the KR approach is superior regarding two other sources of non-convexity, related to the treatment of the amplitude information and the low-frequency information in the data.

\subsection{A Marmousi synthetic test to illustrate the link between convexity and the treatment of the amplitude and low-frequency informations in the data}

\begin{figure}[!ht]
\centering
\includegraphics[width=0.8\linewidth]{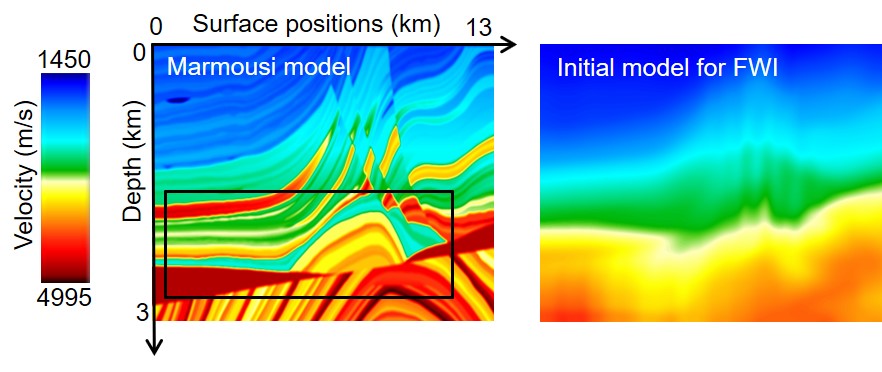}
\caption{
Marmousi 2 model  \cite{Martin_2006_M2E} and initial model for FWI.
}
\label{fig:marm_model}
\end{figure}

\begin{figure}[!ht]
\centering
\includegraphics[width=1.\linewidth]{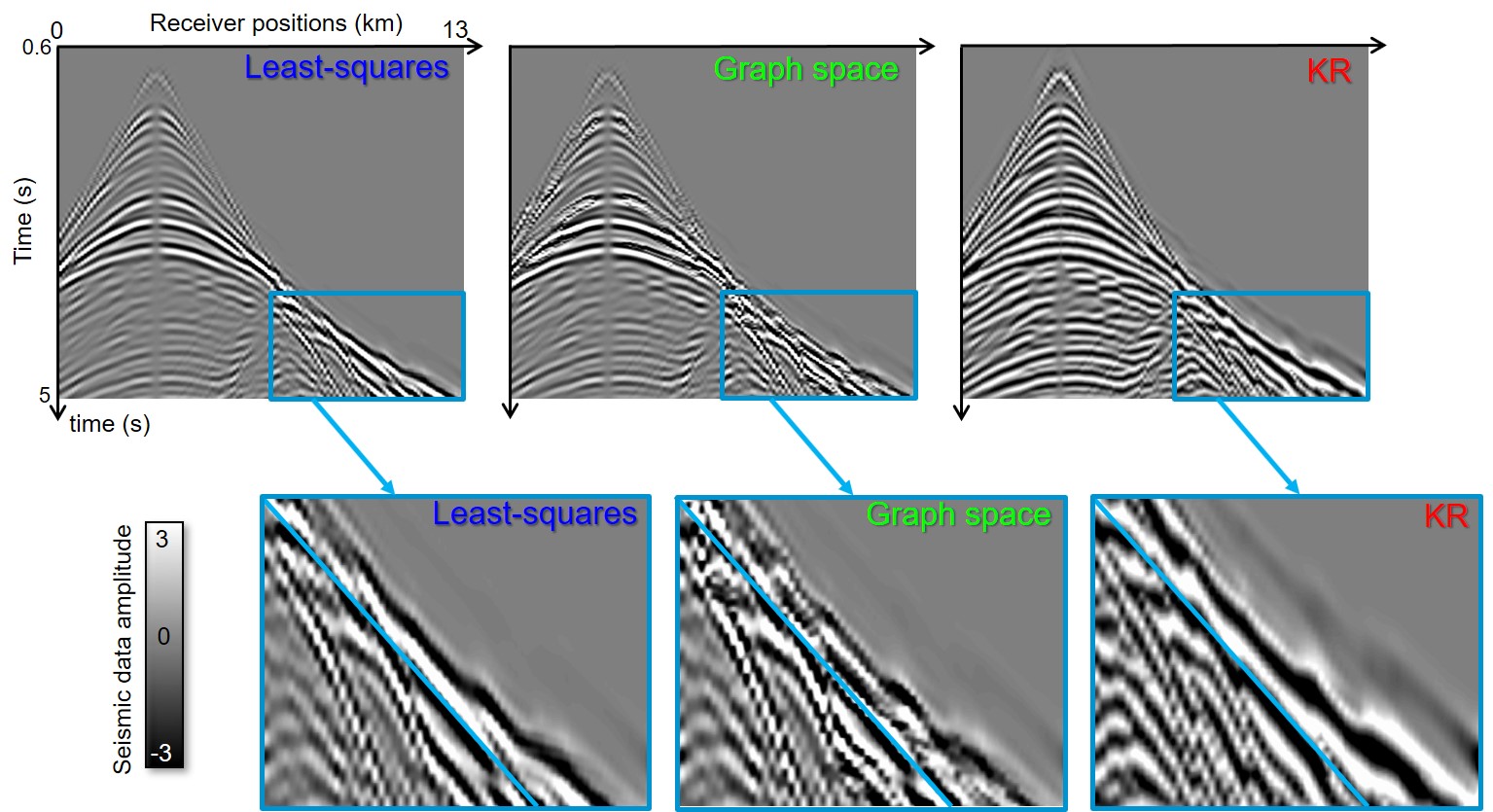}
\caption{
Marmousi 2 data set \cite{Martin_2006_M2E} (Ricker wavelet with peak frequency at 6 Hz and frequencies below 3 Hz muted).
Least-squares, graph-space and KR norm adjoint sources are shown.
}
\label{fig:marm_data}
\end{figure}

We consider a 2D synthetic case called the Marmousi 2 model \cite{Martin_2006_M2E}.
It consists of a 2D velocity model, Figure \ref{fig:marm_model} (left), in which seismic data are modelled with the constant-density acoustic approximation using a Ricker wavelet with peak frequency at 6 Hz that has been low-cut filtered below 3 Hz to be more realistic.
The obtained data will be considered as the observed data for our FWI problem.
The velocity model in Figure \ref{fig:marm_model} (right) will be the initial model for a FWI (it was obtained by Gaussian filtering of the Marmousi 2 model), with the goal being to obtain a model much closer to the Marmousi 2 model at the end of the iterations.
This inverse crime test provides us with a clear benchmark to reach.
Data can be calculated in the initial model in Figure \ref{fig:marm_model} (right) and then corresponding adjoint sources can be computed for various misfit functions.
Figure \ref{fig:marm_data} illustrates the obtained adjoint sources for the least-squares, graph-space and KR norm misfits.\\

Compared to the least-squares adjoint source, the graph-space adjoint source tends to contain shifted events, which is especially visible in the boxes of Figure \ref{fig:marm_data} containing zoomed-in parts of the adjoint sources that contribute the most to FWI.
These shifts or changes in events kinematics explain the improved graph-space convexity with respect to time-shifts.\\

\begin{figure}[!ht]
\centering
\includegraphics[width=0.6\linewidth]{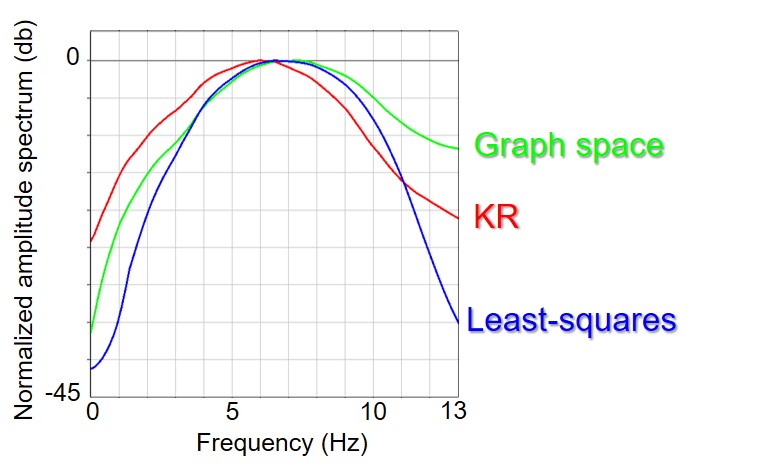}
\caption{
Frequency spectra of the adjoint sources in Figure \ref{fig:marm_data}.
}
\label{fig:marm_data_spectra}
\end{figure}

\begin{figure}[!ht]
\centering
\includegraphics[width=0.55\linewidth]{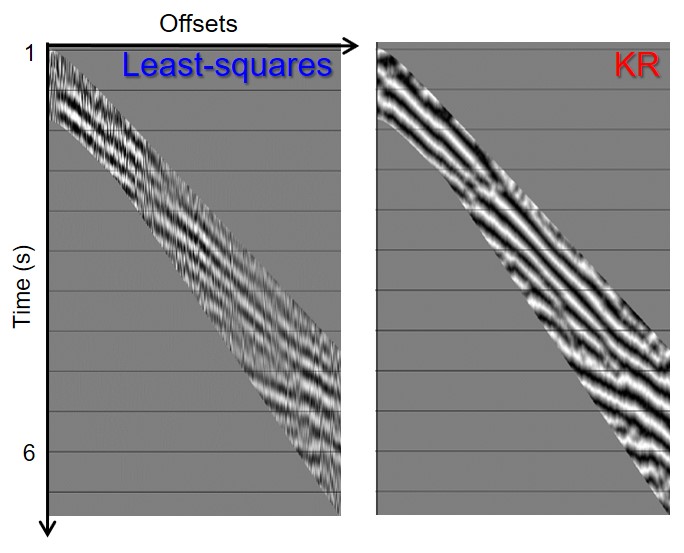}
\caption{
Marine field data at 4 Hz with a mute applied. Least-squares and KR norm adjoint sources are shown.
}
\label{fig:marine_data}
\end{figure}

The KR adjoint source features are very different. 
Firstly, there is an amplitude equalization compared to the least-squares adjoint source, visible in Figure \ref{fig:marm_data}.
This tends to help putting more weight on the events times (or phases) within the FWI, reducing the non-convexity issue.
Secondly, contrariwise to the graph-space adjoint source, there is no change in the events positions but rather a change in the events wavelets. The wavelets become more spread and with a lower frequency content as highlighted in Figure \ref{fig:marm_data_spectra} (note that the higher frequency content of the graph-space adjoint source occurs only because a permutation is not a smooth transform).
The lower frequency content explains the better convexity of KR with respect to time-shifts, compared to least-squares, with the limitation that has been underlined in previous section.
A specificity of the use of the KR norm is to be able to denoise the low frequencies present in the data to some extent and thus to exploit even some very low frequencies (non-exploitable by other methods) to reduce the non-convexity issue.
This can be particularly interesting with noisy field data acquisitions where the quality of the low frequencies in the data could be bad.
Figure \ref{fig:marine_data} shows how the KR adjoint source compares to the least-squares adjoint source for marine field data with a mute applied (common in industrial FWI situations). 
The noise in the data is strong  and differs trace-to-trace, degrading the continuity of the least-squares adjoint source. Interestingly, the KR adjoint source is strongly denoised, with an increased continuity in the direction of the events and better amplitude balancing.
This may be useful to start FWI at an even lower frequency to mitigate the non-convexity.\\

\begin{figure}[!ht]
\centering
\includegraphics[width=1.1\linewidth]{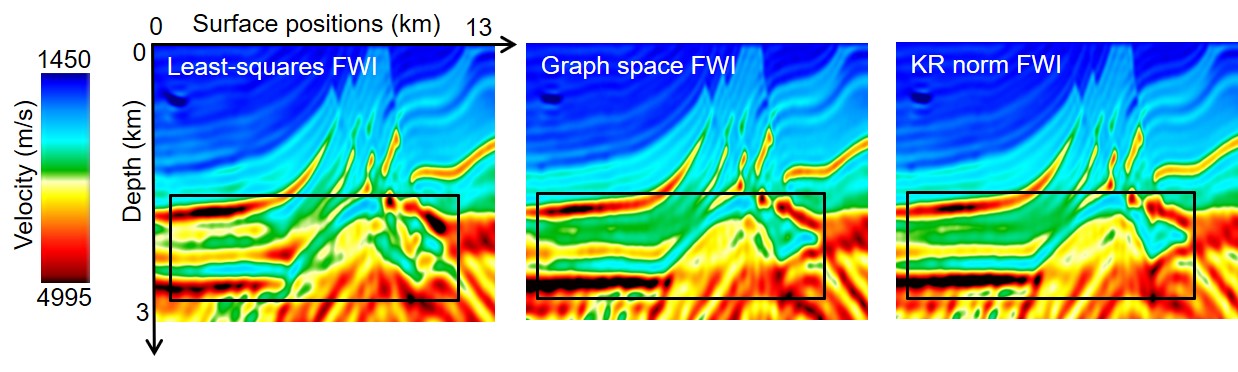}
\caption{
Marmousi 2 model  \cite{Martin_2006_M2E}.
FWI inversion performing 20 iterations directly at 10 Hz.
}
\label{fig:marm_fwi}
\end{figure}

As we can see,  graph-space and KR FWI each have their strengths, which are related to complementary features.
After the adjoint sources analysis, we study if one of these methods gives better FWI results.
Using the Marmousi configuration, we start the FWI from the smooth initial velocity model in Figure \ref{fig:marm_fwi} (right), performing 20 iterations  directly at up to 10 Hz, .
The models estimated by graph-space and KR FWI match the Marmousi 2 model in Figure \ref{fig:marm_model} (left) much better than the models estimated by least-squares FWI. This is especially true in the highlighted zones where the poor least-squares result can be related to non-convexity issues, i.e., least-squares FWI is stuck in a local minimum.
Interestingly, we did not find Marmousi 2 configurations where graph-space FWI outperformed KR FWI or vice versa.
It seems that both graph-space and KR FWI manage to mitigate the non-convexity issues to a similar level in the Marmousi 2 case, while working very differently on the data (shifting events for graph-space, and enhancing the amplitudes balancing, low frequencies and events continuity for KR).

\newpage
\subsection{Industrial applications of the Kantorovich-Rubinstein strategy to various 3D field data}
\label{sec:KR_industrial}

Many successful industrial applications of KR FWI on 3D field data have been published, see
for instance \cite{Poncet2018,Messud2019,Sedova2019,Hermant2019,Carotti2020,Hermant2020}.
In this Section, we review three examples. For further details or more illustrations, the reader is invited to refer to the aforementioned articles.

\subsubsection{North of Oman land data}

\begin{figure}[!ht]
\centering
\includegraphics[width=0.83\linewidth]{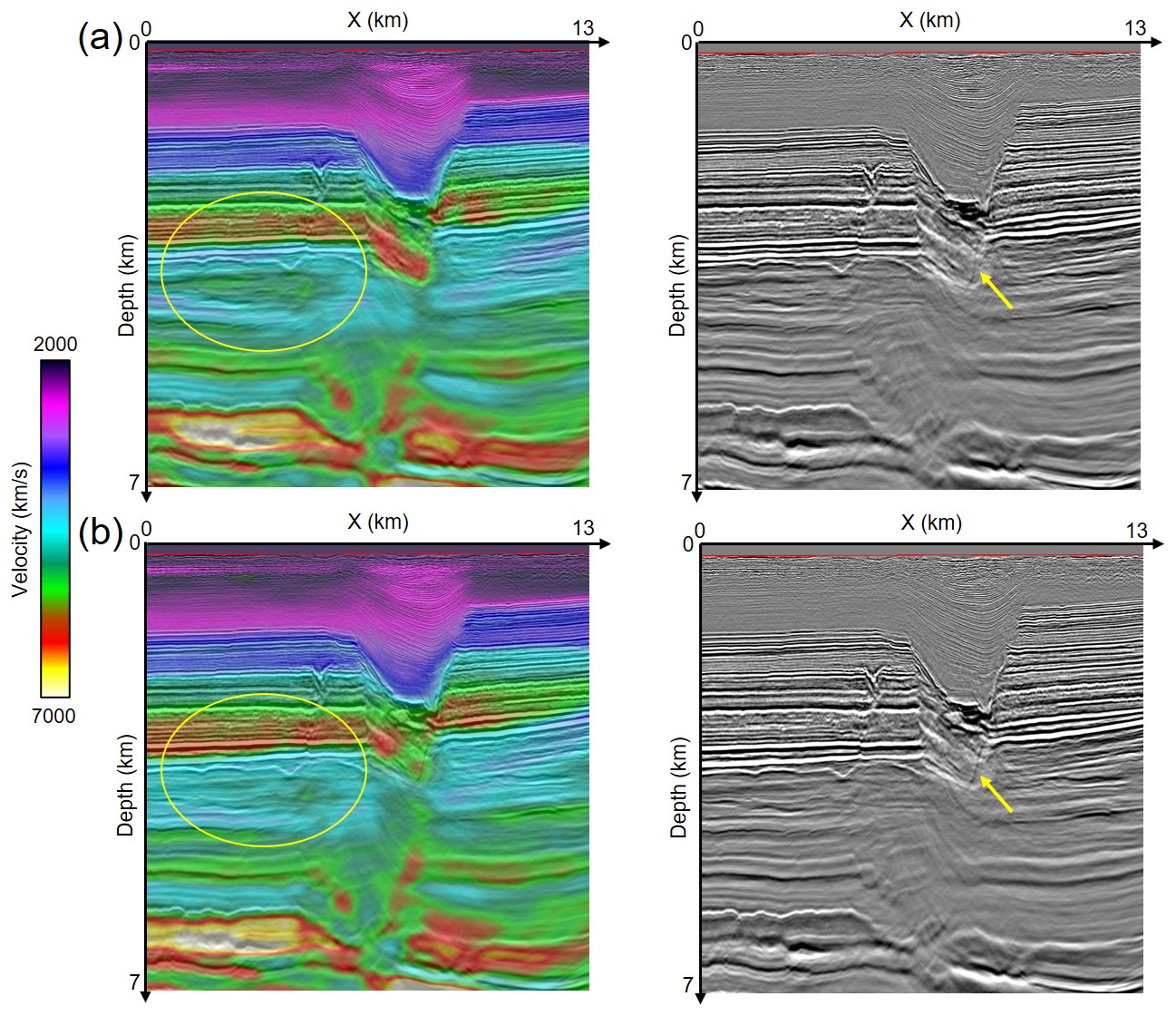}
\caption{
North of Oman data. 
(a) Least-squares and (b) KR FWI results at 16 Hz
(obtained in the same configuration).
Left: velocity model inverted by FWI superimposed on images of the subsurface reflectors (obtained using the FWI model into a ``Kirchhoff'' depth migration algorithm).
The ovals highlight the improved velocity contrast 
and the correction of the velocity increase achieved by KR FWI.
Right: images of the subsurface reflectors alone;
the arrows highlight the improved focusing of a fault achieved by KR FWI.
From \cite{Messud2021}.
}
\label{fig:figure6}
\end{figure}

The first example refers to 3D land data
%with full-azimuth and offsets of up to 13 km
acquired in the North of Oman (see \cite{Carotti2020} for more details).
The FWI was run with the frequency increasing from 2 Hz to 16 Hz, using a pseudo-acoustic wave propagation and following the data pre-processing workflow proposed by 
\cite{Sedova2019}.
%Starting from a heavily smoothed VTI initial model obtained using a previous FWI result. 
Figure \ref{fig:figure6} compares the least-squares and KR FWI results for a subsurface velocity inversion.
%using the same input data and starting model.
The oval in Figure \ref{fig:figure6} highlights the improved velocity contrast achieved by KR FWI, along with the correction of an unexpected velocity increase produced by least-squares FWI.\\

The FWI velocity models can be further used in a ``depth migration'' algorithm, whose aim is to provide an images of the subsurface  ``reflectors'' (oor discontinuities).
Details on such a method can be found in \cite{Bleistein_1987_IRE,Aki_1980_QST,Claerbout_1985_IEI}.
Such subsurface reflectors images have been superimposed to their corresponding FWI velocity models in Figure \ref{fig:figure6} (left)
and are shown alone in Figure \ref{fig:figure6} (right).
We can observe that the KR FWI velocity model provides a better or more focused
image of the subsurface deep reflectors than the least-squares FWI model, especially at the position of a major fault as highlighted by the yellow arrows.
This contributes to demonstrate the superiority of KR FWI over least-squares FWI\\

This example illustrates, in a challenging land acquisition context, how the better convexity properties of the KR norm translate into a better FWI model.

\subsubsection{North Sea marine data}

\begin{figure}[!ht]
\centering
\includegraphics[width=1.14\linewidth]{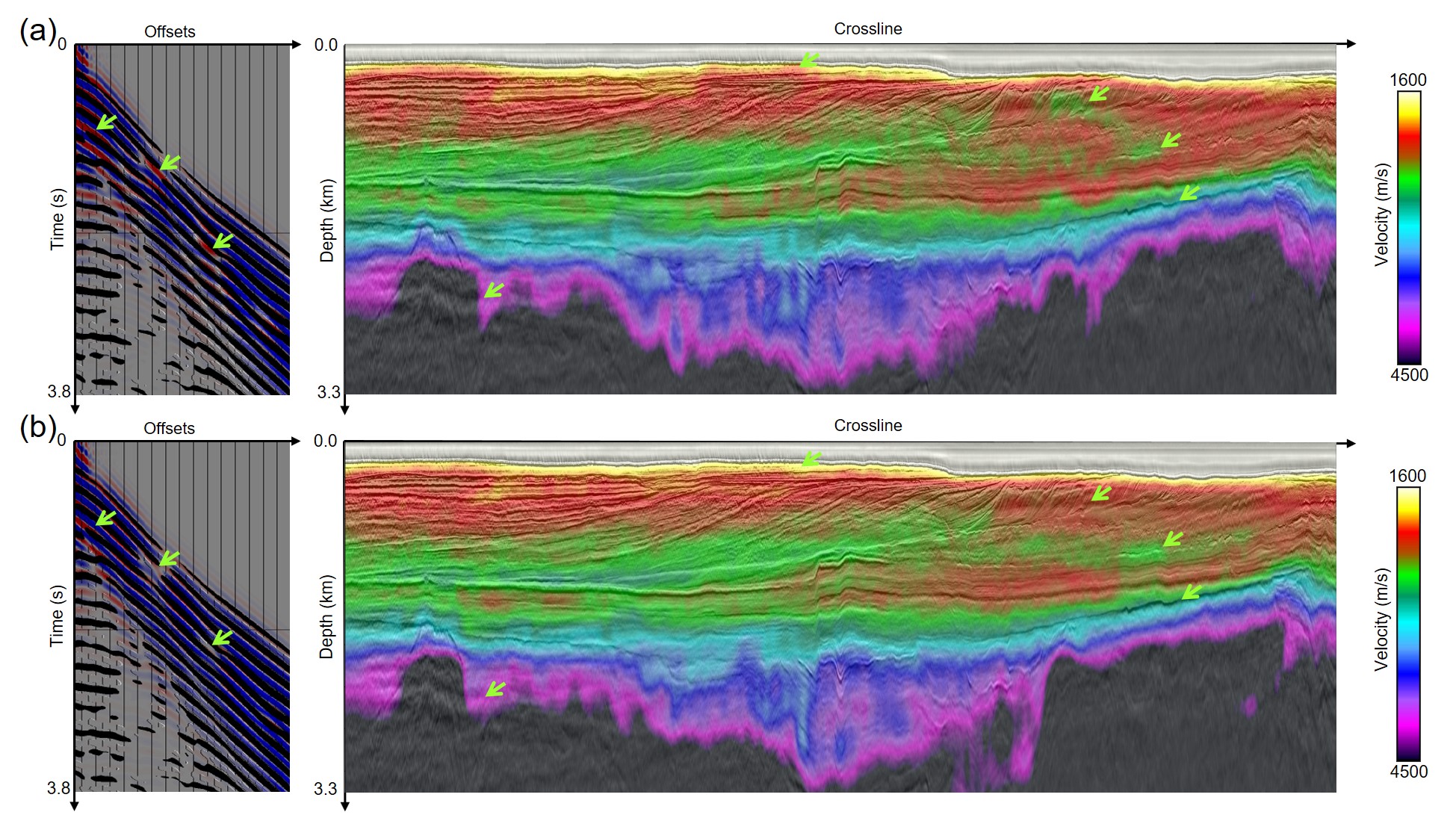}
\caption{
North Sea data.
(a) Least-squares and (b) KR FWI results at 7 Hz
(obtained in the same configuration).
Left: observed data (in black-grey-white) superimposed on the data calculated in corresponding FWI-updated model (red-blue)
(the arrows highlight where the calculated data suddenly jumps from one cycle to another when using the least-squares model).
Right: FWI-updated models superimposed on corresponding images of the subsurface reflectors
(the arrows highlight areas where KR FWI gives an improved velocity model).
From \cite{Messud2021}.
}
\label{fig:figure7}
\end{figure}

The second example refers to North Sea marine data (see \cite{Messud2019} for more details).
Figure \ref{fig:figure7}
shows results obtained with a 7 Hz FWI inversion of the subsurface velocity.
%  using the same input data and starting model.
Figure \ref{fig:figure7} (left) shows the observed data overlaid on top of the data calculated in the FWI updated model.
At the position of the green arrows, we can observe that least-squares FWI leads to ``red spots''.
These red spots are due to events that suddenty jump from one ``cycle'' to another in the calculated data, a typical cause of cycle-skipping, which allows to deduce that least-squares FWI get stuck in a local minimum.
Contrariwise, KR FWI does not exhibit red spots, an indication of an absence of cycle-skipping.\\

Figure~\ref{fig:figure7} (right) shows the least-squares and KR FWI models superimposed on corresponding images of the subsurface reflectors.
We can observe that the least-squares FWI model does not ``follow'' the structures in the the subsurface reflectors images, i.e.,
it lacks structural consistency, especially in the zones highlighted by green arrows. This illustrates how the cycle-skipping can affect the FWI result.
Conversely, the KR FWI model exhibits a better structural consistency, inverting for an improved velocity model.

\subsubsection{Barents Sea marine data}

\begin{figure}[!ht]
\centering
\includegraphics[width=1.\linewidth]{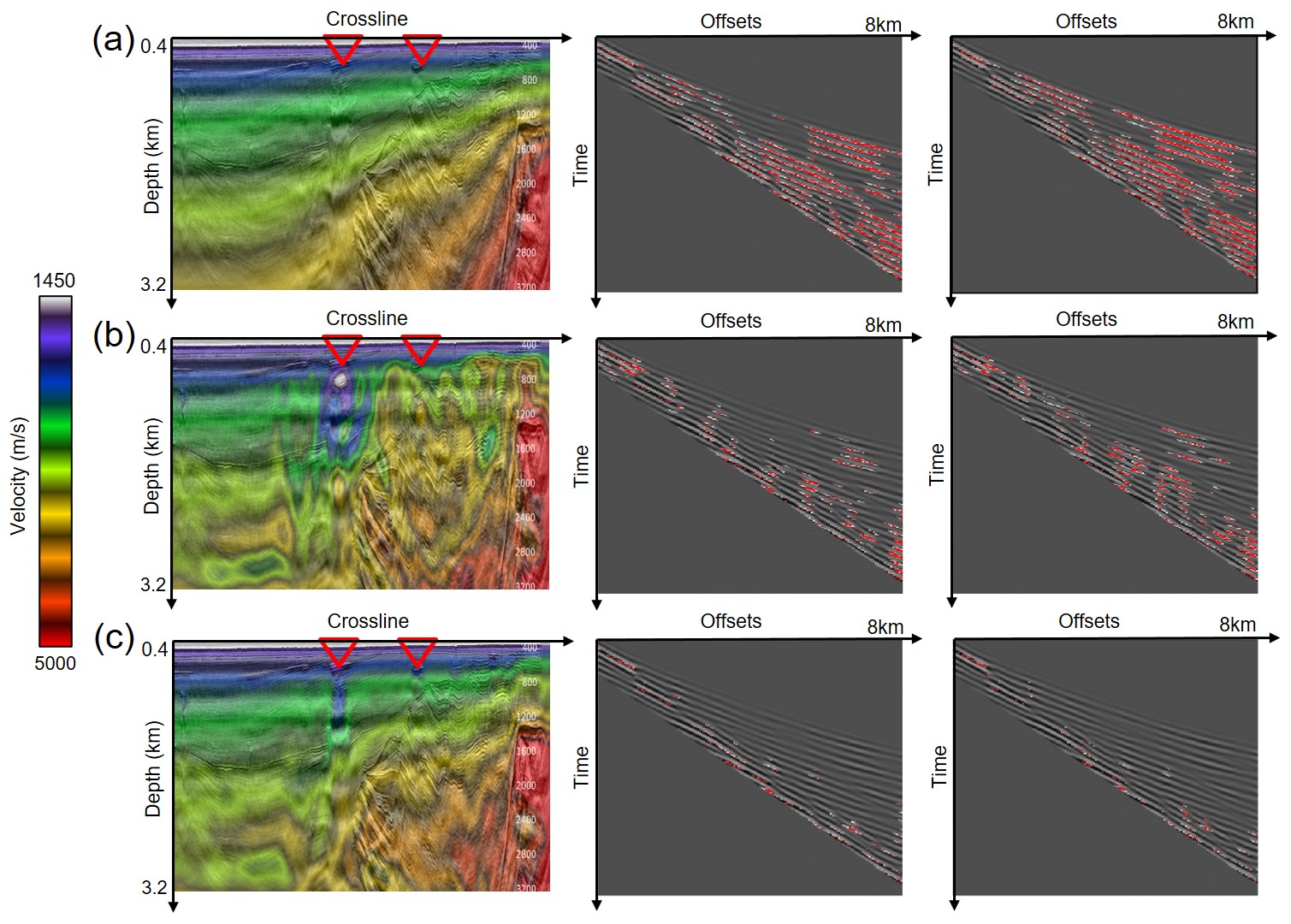}
\caption{
Barents Sea data.
(a) Initial FWI model, (b) least-squares FWI result and (c) KR FWI result at 6 Hz
(obtained in the same configuration).
%Results of velocity model building using LSQ FWI (left) and multiD KR FWI (right) at 6 Hz.
Left: FWI updated models superimposed on corresponding images of the subsurface reflectors.
Middle and right: normalized absolute values of the difference between observed data and data calculated in FWI-updated models
(red means large values and thus poor matching)
at the positions highlighted by the red triangles in the left figures.
From \cite{Messud2021}.
}
\label{fig:figure8}
\end{figure}

The last example refers to Barents Sea marine data (see \cite{Carotti2020} for more details).
It is challenging because of gas accumulations of varying sizes and depths that can lead to instabilities within FWI.
The poor initial model in Figure \ref{fig:figure8} (top-left) has been used to initialize the FWI iterations.
%KR FWI managed to converge to a structurally consistent model, while better matching the observed data.
Figure \ref{fig:figure8} (left)
illustrates the improved FWI-updated model obtained using KR compared to least-squares.
With corresponding images of the subsurface reflectors superimposed, we observe that KR FWI leads to more structural consistency and less instabilities, an indication that the inverted velocity model is better.\\

Figure \ref{fig:figure8} (middle and right) shows the normalized absolute value of the difference between observed data and data calculated in the FWI models, less red indicating a better data matching after FWI.
Of course, both least-squares and KR FWI improve the data matching compared to the one related to the initial model (by inverse problem construction).
However, KR FWI shows a much better matching than least-squares FWI,
which is an indication that KR FWI has converged to a better minimum thanks its enhanced convexity.\\

\subsection{Application of the graph-space strategy to 3D field data: the Valhall case study}

The results presented here are extracted from a recently submitted study \cite{Pladys_2021_OTV}.

\subsubsection{Data acquisition and context}

The Valhall field is located in the southern part of the Norwegian sector in the North Sea, approximately $300$~km southwest of Stavanger. This field was discovered in 1975 and it has been used since then for oil production. An oil reservoir is located below trapped gas in tertiary shales. This trapped gas forms a low-velocity zone acting as a screen, making imaging below it challenging.\\ 

\begin{figure}
   \centering
   \includegraphics[width=1\linewidth]{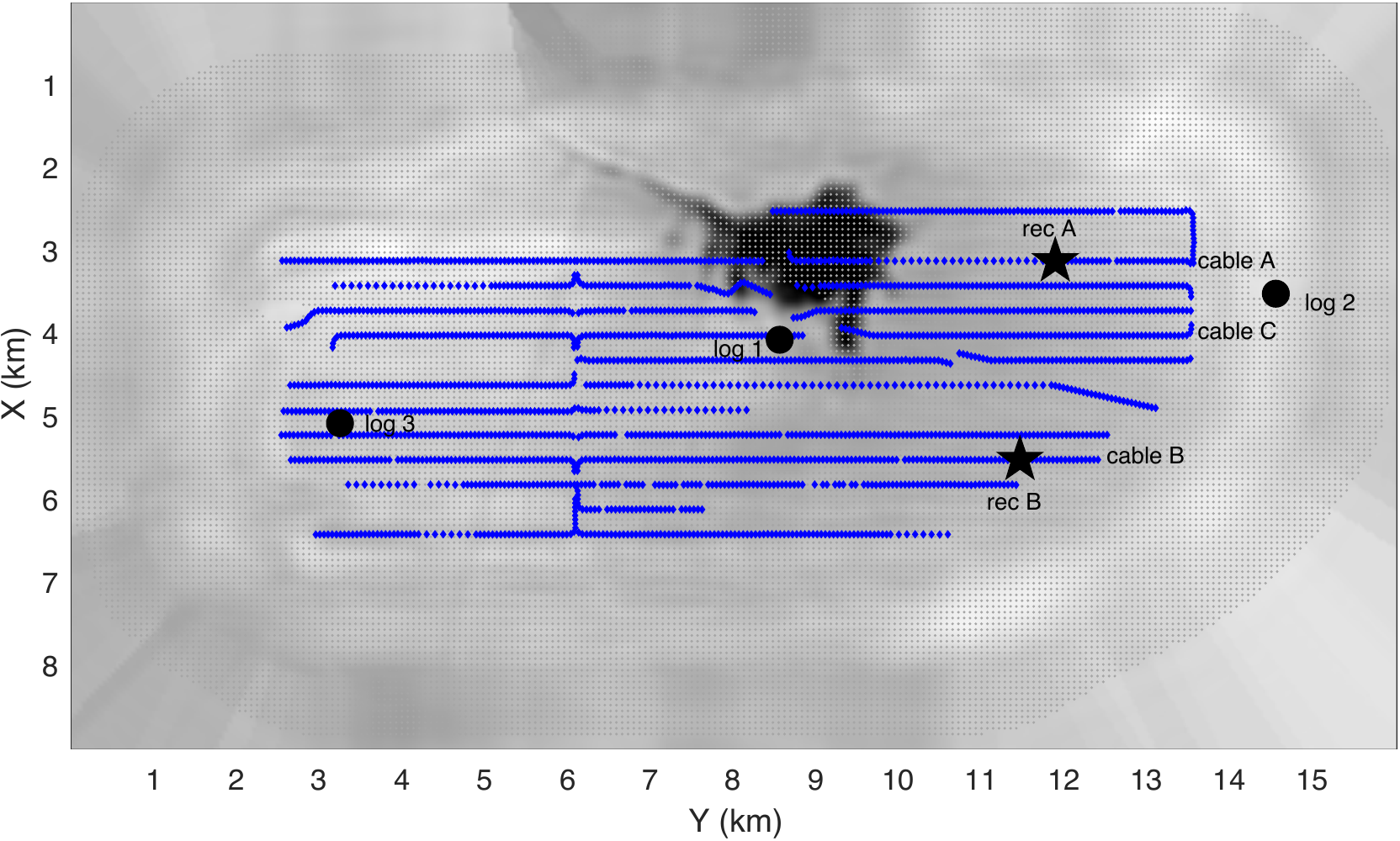}
   \caption{Layout of the Valhall acquisition overlapped on an horizontal P-wave velocity slice at 1~km obtained by FWI. Location of sources (gray dots) and receivers (blue diamonds). Two receivers positions (A and B) are located with black stars. Cables A ($x=2950$~m), B ($x=5530$~m) and C ($x=3080$~m) are identified.}
   \label{valhall:acqui}
\end{figure}

Thanks to a shallow-water environment (the water depth is approximately $70$ m), the deployment of ocean-bottom cables (OBC) with 4-component receivers (hydrophones measuring the pressure + 3 components geophones measuring the displacement) was relatively  easy. Twelve receiver cables were deployed on the seabed, containing $2048$ receivers with an inline spacing of $50$ m and a cable spacing of $300$ m. On the surface, a total of $50824$ shots of pressure airgun sources were performed, $5$ m below the surface. The layout of this 3D acquisition is presented in Figure \ref{valhall:acqui}. The imaged zone represents a volume of $9 \times 16 \times 4.5 $ km$^{3}$, discretized on a $50$ m Cartesian grid at the finest level, leading to $181 \times 321 \times 91$ discrete unknowns. In this study, we use only the hydrophone component of the acquisition performed in 2011 \cite{Barkved_2003_LoFS}. This 3D dataset was made available to us thanks to AkerBP, one of the companies that supports the SEISCOPE project.\\

We have investigated the Valhall case study quite extensively over the past few years \cite{Prieux_2011_FAI,Prieux_2013_MFWa,Operto_2015_ETF,Kamath_2020_MFW}, with successful results based on the conventional tomography + multi-scale approach described in the introduction. This makes the Valhall case study an adequate playground for testing new FWI methodologies such as the use of optimal transport distances.\\

To highlight the interest for using the graph-space strategy, we present the results obtained when starting from two different initial velocity models. The first is accurate and is obtained through reflection tomography. It has been provided to us by AkerBP. Starting from this initial model, and interpreting the data in a multiscale manner using the two frequency bands $2.5 - 5 $ Hz and $2.5 - 7 $ Hz, least-squares based FWI converges towards a plausible 3D velocity model which satisfactorily explains the data. The second initial model is purposely rough, varying only along the vertical direction. It generates important time-delays in the waveform, which in turn prevents least-squares FWI from converging towards a correct estimation of the subsurface velocity due to cycle-skipping. We show how the use of the graph-space optimal transport strategy can help mitigate this effect.\\

For this field data application, we rely on a 3D visco-acoustic anisotropic approximation of the wave propagation. Taking into account both attenuation and anisotropy has shown to be important to correctly interpreting the data, while elastic propagation effects can be neglected as their imprint is weak on the hydrophone component of the data at this frequency range. In this frame, the subsurface is described by the P-wave velocity, attenuation and density models, and anisotropy models related to a vertical transverse isotropy approximation of the subsurface anisotropy (basically, the vertical velocity differs from the horizontal velocity as an effect of thin sub-wavelength horizontal layering of the subsurface). More details on the implementation of our 3D visco-acoustic anisotropic modeling and inversion methods can be found in \cite{Yang_2018_TRN}. We invert only for the P-wave velocity models, and consider the other models as fixed. They are determined prior to the inversion from different means: the density is inferred from the initial P-wave velocity model using Gardner's law, the attenuation is considered as homogeneous below the water layer, and the anisotropy models are obtained from reflection tomography. They have also been provided to us by AkerBP.

\subsubsection{FWI starting from an accurate initial model from reflection tomography}

\begin{figure}[!ht]
   \centering
   \includegraphics[width=0.8\linewidth]{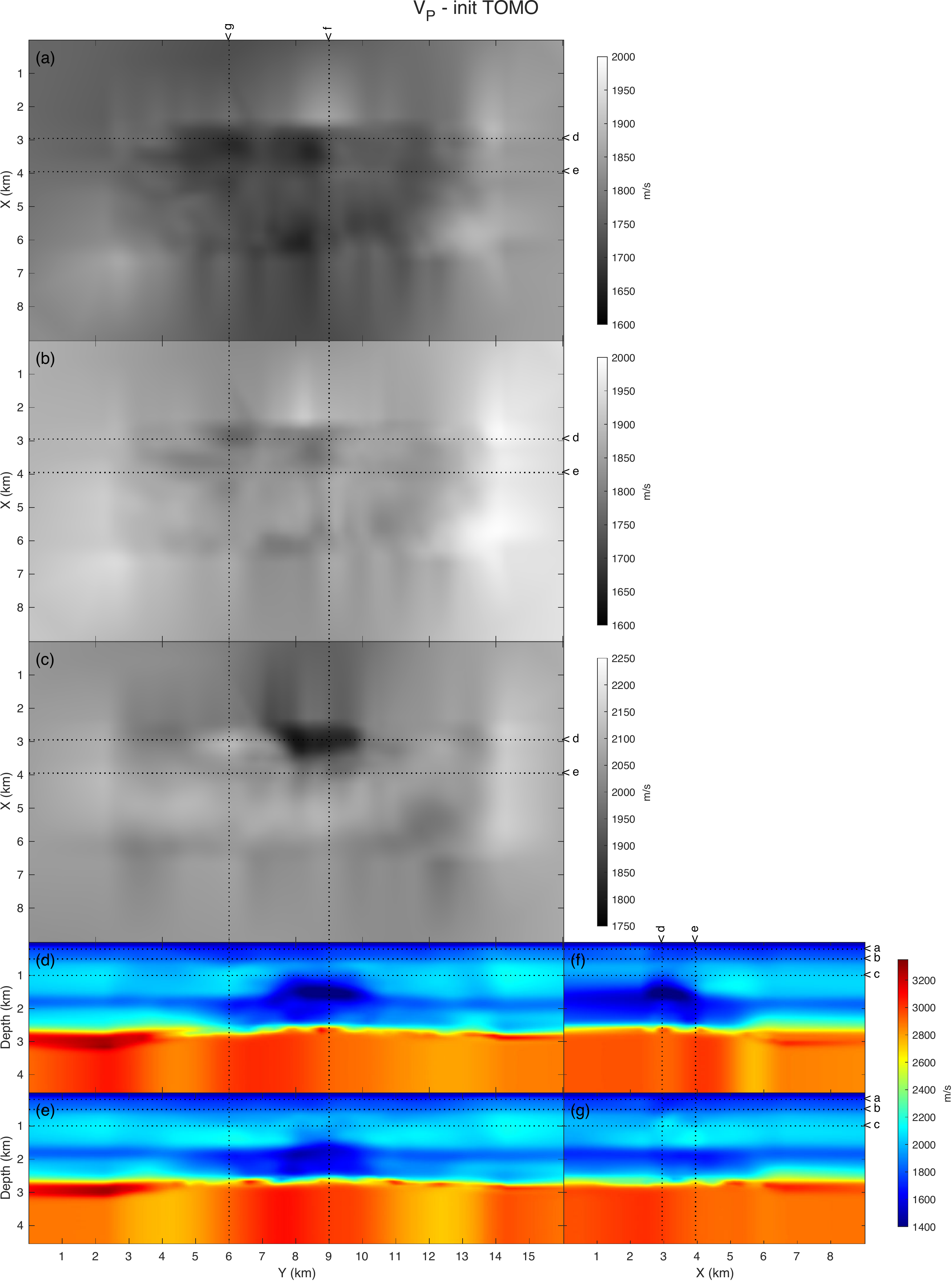}
   \caption{Slices of the initial tomography model. (a-c) Horizontal slices at (a) $0.2$ km depth, (b) $0.5$ km depth and (c) $1$~km depth. (d-e) Inline vertical slices for (d) $x=2.95$~km and (e) $x=3.95$~km. (f-g) Cross-line vertical slices at (f) $y=9$~km and (g) $y=6$~km.}
   \label{init:VP_initTOMO}
\end{figure}

The accurate initial model obtained by reflection tomography is presented in Figure \ref{init:VP_initTOMO}. In the different horizontal (gray scale) and vertical (color scale) slices, we can recognize a central low-velocity anomaly corresponding to the presence of trapped gas in the sediment layers. The horizontal slices are extracted at relatively shallow depths (0.2 km, 0.5 km, and 1 km), while the depth slices give a view of the velocity model down to 4.5 km. Below this low velocity anomaly appears a strong interface corresponding to a harder rock zone (constituted of chalk). This is the top of the reservoir, which is located below this interface. As can be seen, this initial tomography model is ``blurred'': no detailed information can be recovered or directly interpreted from it. 

% \begin{figure*}
%     \centering
%     \includegraphics[width=0.75\linewidth]{eta}
%     \caption{Same as \ref{init:VP_initTOMO} for anisotropic parameter $\eta$.}
%     \label{init:eta}
% \end{figure*}

\begin{figure}[!ht]
   \centering
   \includegraphics[width=0.8\linewidth]{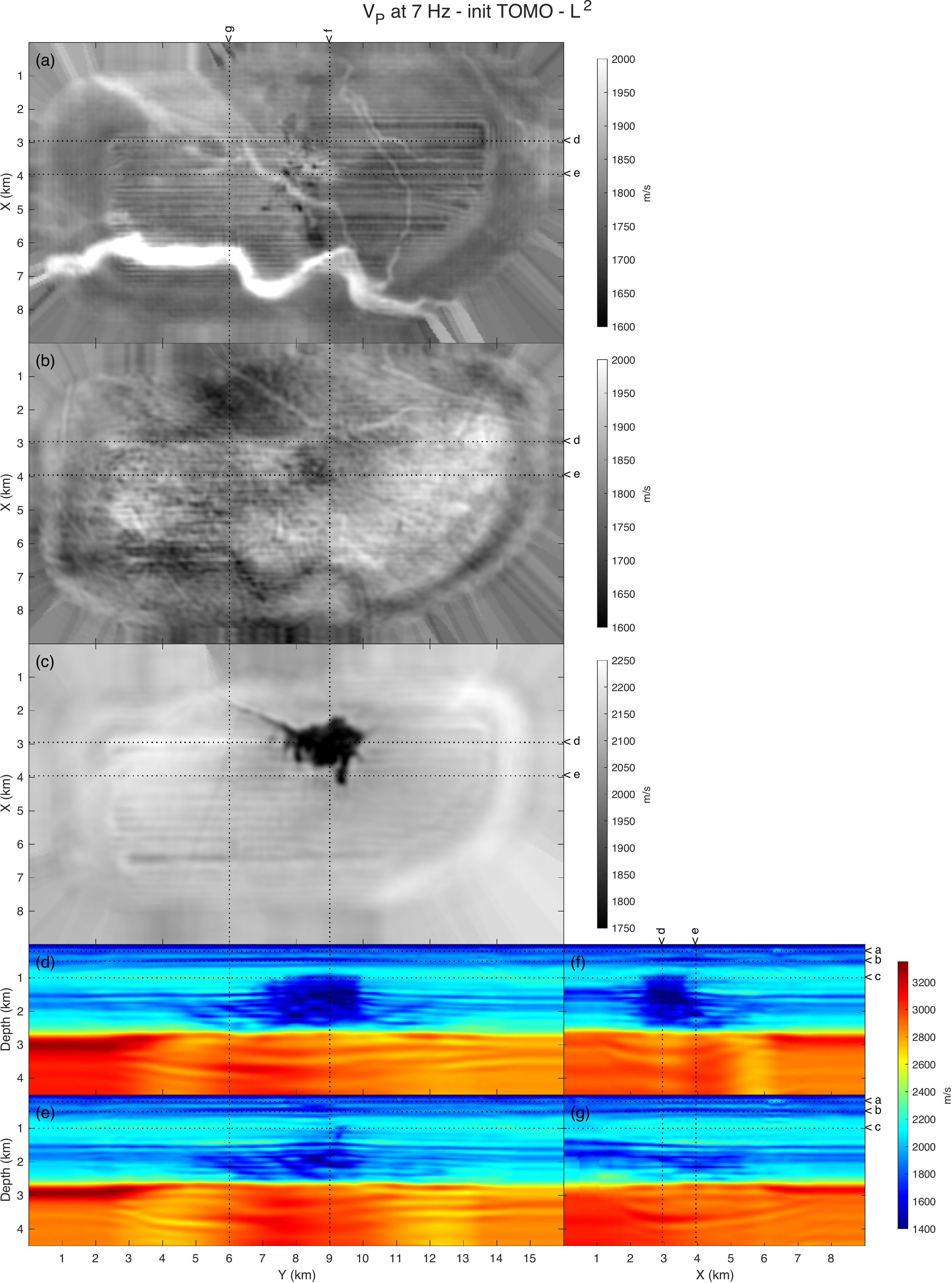}
   \caption{Slices of the $7$~Hz FWI reconstructed velocity using a least-squares misfit function starting from the initial tomography model. (a-c) Horizontal slices at (a) $0.2$ km depth, (b) $0.5$ km depth and (c) $1$ km depth. (d-e) Inline vertical slices for (d) $x=2.95$~km and (e) $x=3.95$~km. (f-g) Cross-line vertical slices at (f) $y=9$~km and (g) $y=6$~km.}
   \label{vp:VP_initTOMO_7Hz_L2}
\end{figure}

The results obtained using a conventional multi-scale least-squares based FWI are presented in Figure \ref{vp:VP_initTOMO_7Hz_L2}. The resolution improvement is impressive: the delineation of the trapped gas zone is much clearer. Also, at $0.2$ km depth, a network of submarine channels is revealed with great accuracy. At $0.5$ km depth, coherent ``line'' features are interpreted as scrapped on the seabed left by drifting icebergs. This result, in agreement with previous 3D investigations \cite{Sirgue_2010_FWI,Operto_2015_ETF}, is a clear illustration of the resolution power of FWI when sufficiently low-frequency data and sufficiently accurate initial models are available. 

% \begin{figure*}
%     \centering
%     \includegraphics[width=1\linewidth]{log_initTOMO_7Hz}
%     \caption{Comparison of \vp profiles extracted from the \moda initial model (dashed red), $7$~Hz FWI models using \ls (solid yellow) and $7$~Hz FWI model using GSOT (solid purple) with sonic log (solid black). Left subfigure corresponds to the Log 1 at the center of the target. Middle subfigure to the Log 2, and right subfigure to Log 3 (which is far away from the target).}
%     \label{vp:log_initTOMO_7Hz}
% \end{figure*}

\begin{figure}[!ht]
   \centering
   \includegraphics[width=1\linewidth]{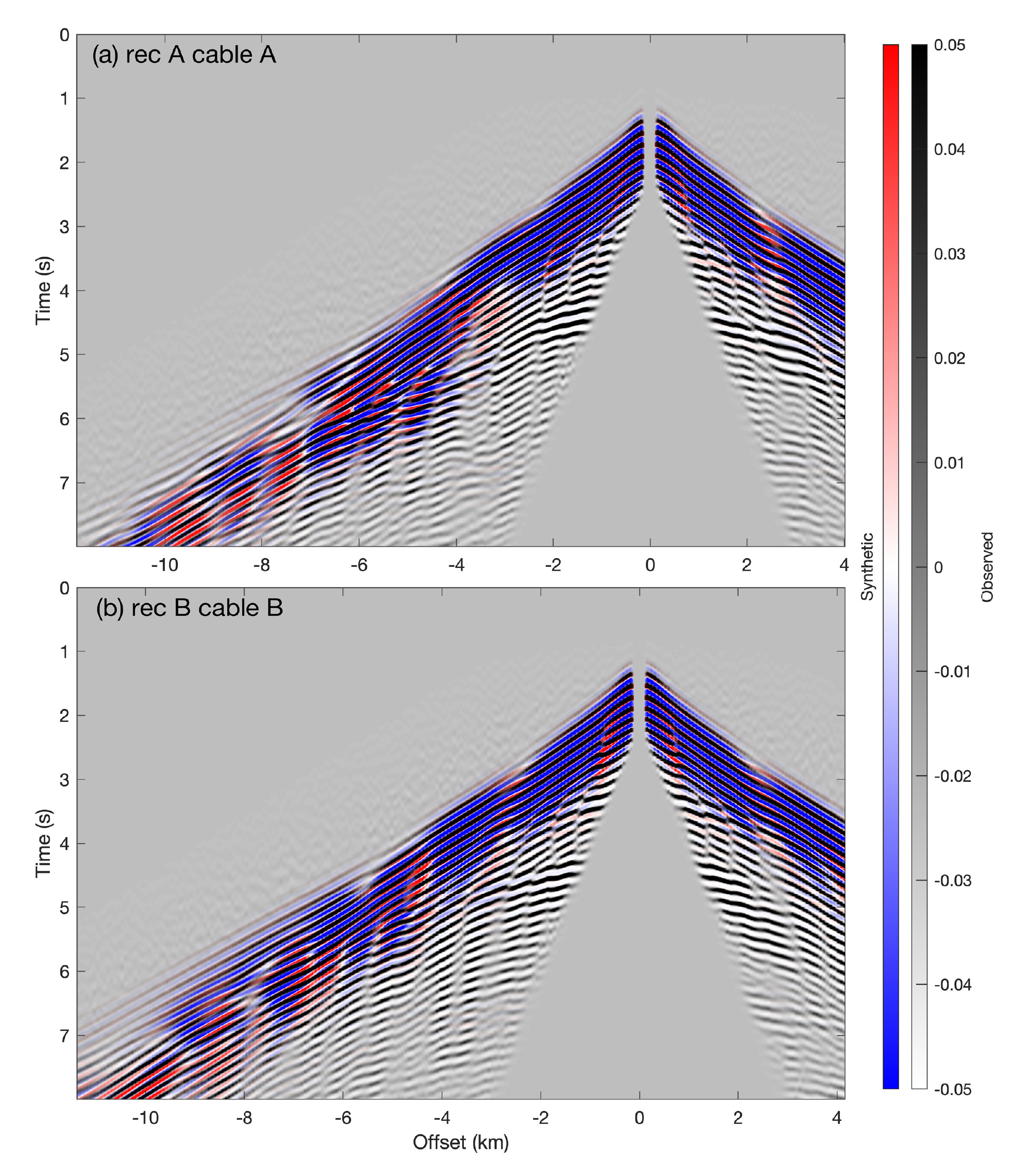}
   \caption{2D common-receiver gathers at $7$~Hz starting from the initial tomography model. Synthetic data (blue/white/red color scale) generated into the final reconstructed velocity model using the least-squares misfit function. (a) receivers along cable A (through the low velocity anomaly). (b) receiver B along cable B. Field data are overlapped in gray-scale with transparency.}
   \label{crg:CRG_7Hz_initTOMO}
\end{figure}

The analysis of the fit to the data (comparison between calculated and field data) in the final model is given in Figure \ref{crg:CRG_7Hz_initTOMO}. The calculated data is presented in color scale, while the field data is overlapped in gray scale with transparency to analyze the match between the datasets. The central part of the data has been muted as it contains the imprint of Sch\"olte waves, propagating at the fluid/solid interface, which cannot be predicted in the acoustic approximation we are using in this experiment. A good match between the calculated and observed data can be observed. 

\subsubsection{FWI starting from a rough initial model: interest of the graph-space OT approach}

\begin{figure}[!ht]
   \centering
   \includegraphics[width=0.8\linewidth]{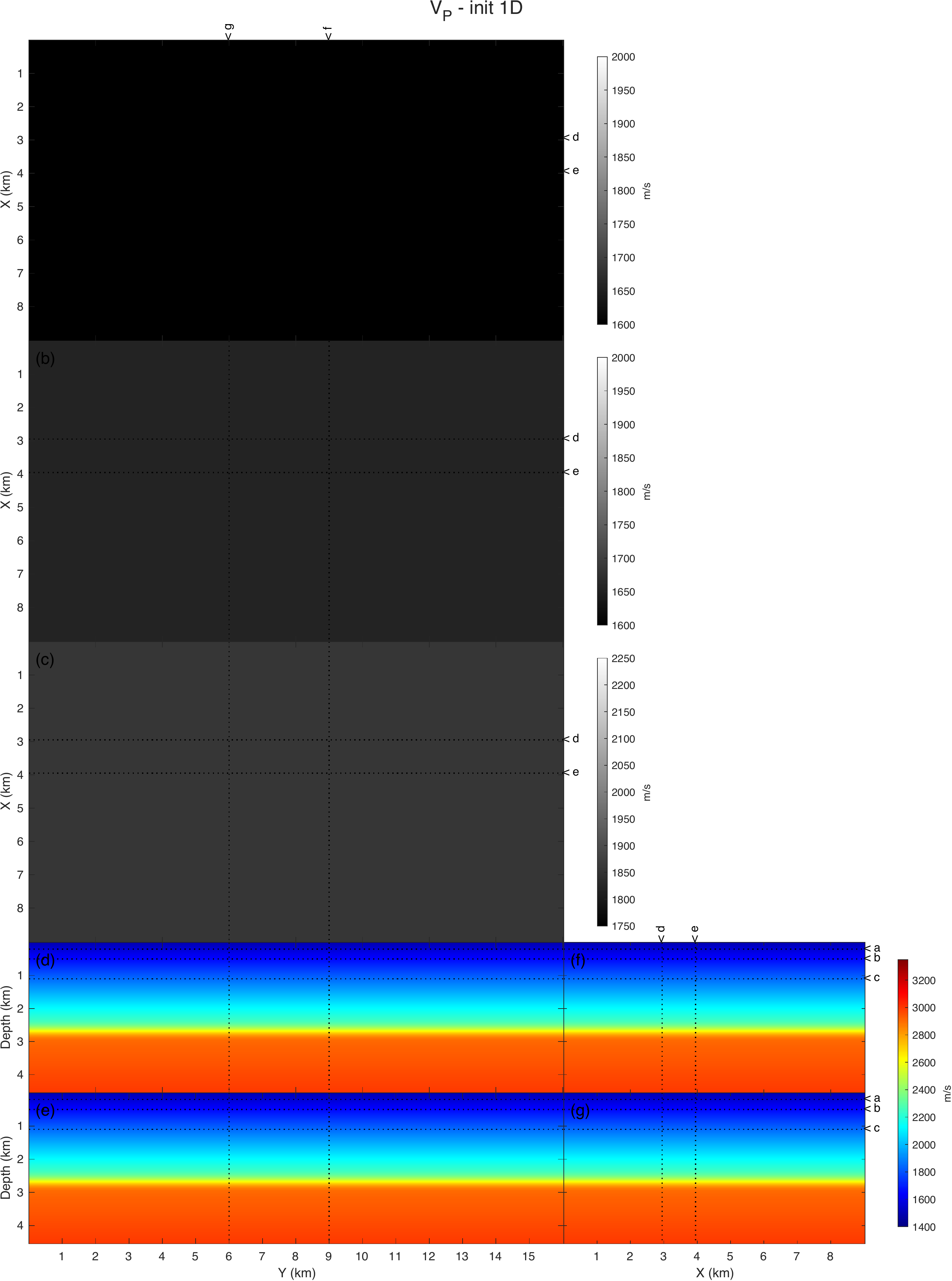}
   \caption{Slices of the rough initial model. (a-c) Horizontal slices at (a) $0.2$ km depth, (b) $0.5$ km depth and (c) $1$ km depth. (d-e) Inline vertical slices for (d) $x=2.95$~km and (e) $x=3.95$~km. (f-g) Cross-line vertical slices at (f) $y=9$~km and (g) $y=6$~km.}
   \label{vp:VP_init1D}
\end{figure}

\begin{figure}[!ht]
   \centering
   \includegraphics[width=0.8\linewidth]{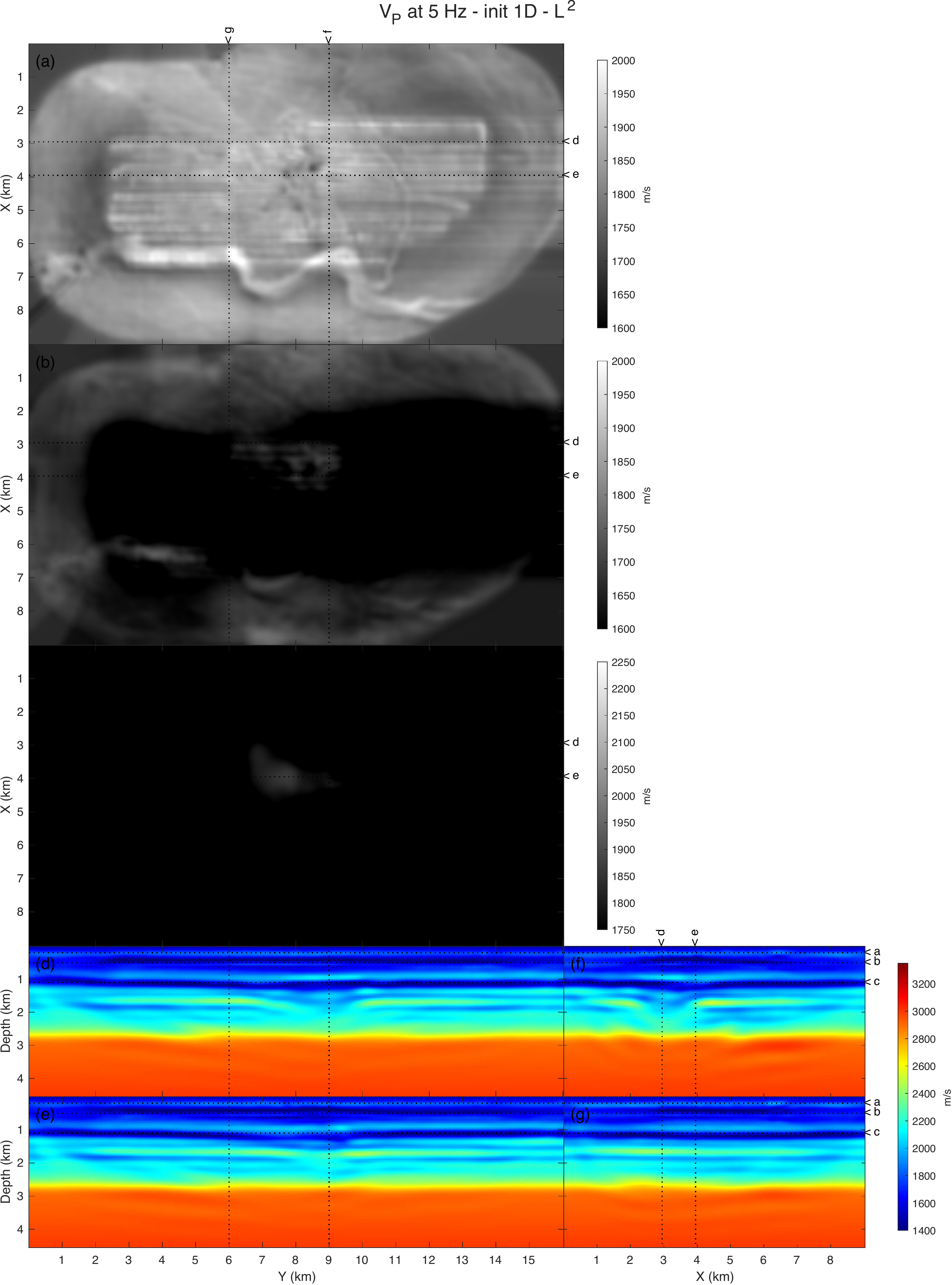}
   \caption{Slices of the $5$ Hz FWI reconstructed velocity model using the least-squares misfit starting from the rough initial model. (a-c) Horizontal slices at (a) $0.2$ km depth, (b) $0.5$ km depth and (c) $1.1$ km depth. (d-e) Inline vertical slices for (d) $x=2.95$~km and (e) $x=3.95$~km. (f-g) Cross-line vertical slices at (f) $y=9$~km and (g) $y=6$~km.}
   \label{vp:VP_init1D_5Hz_L2}
\end{figure}

The rough initial model we consider is presented in Figure \ref{vp:VP_init1D}. As previously explained, this model varies only with depth; therefore, the horizontal slices exhibit constant velocity values. 
The results obtained using a conventional least-squares FWI at the first-frequency band $2.5 - 5$ Hz are presented in Figure \ref{vp:VP_init1D_5Hz_L2}. As can be seen, this FWI was not able to converge towards a meaningful velocity model. Only the shallowest part of the model provides some details about the network of channels identified in Figure \ref{vp:VP_initTOMO_7Hz_L2}, however with an incorrect background velocity. Deeper, the updates of the velocity are performed in the opposite direction of what would be required, which is typical of cycle-skipping. As the least-squares inversion fails already at the first frequency band, we do not continue with the multi-scale workflow and stop the inversion at $5$ Hz.\\ 

\begin{figure}[!ht]
   \centering
   \includegraphics[width=0.8\linewidth]{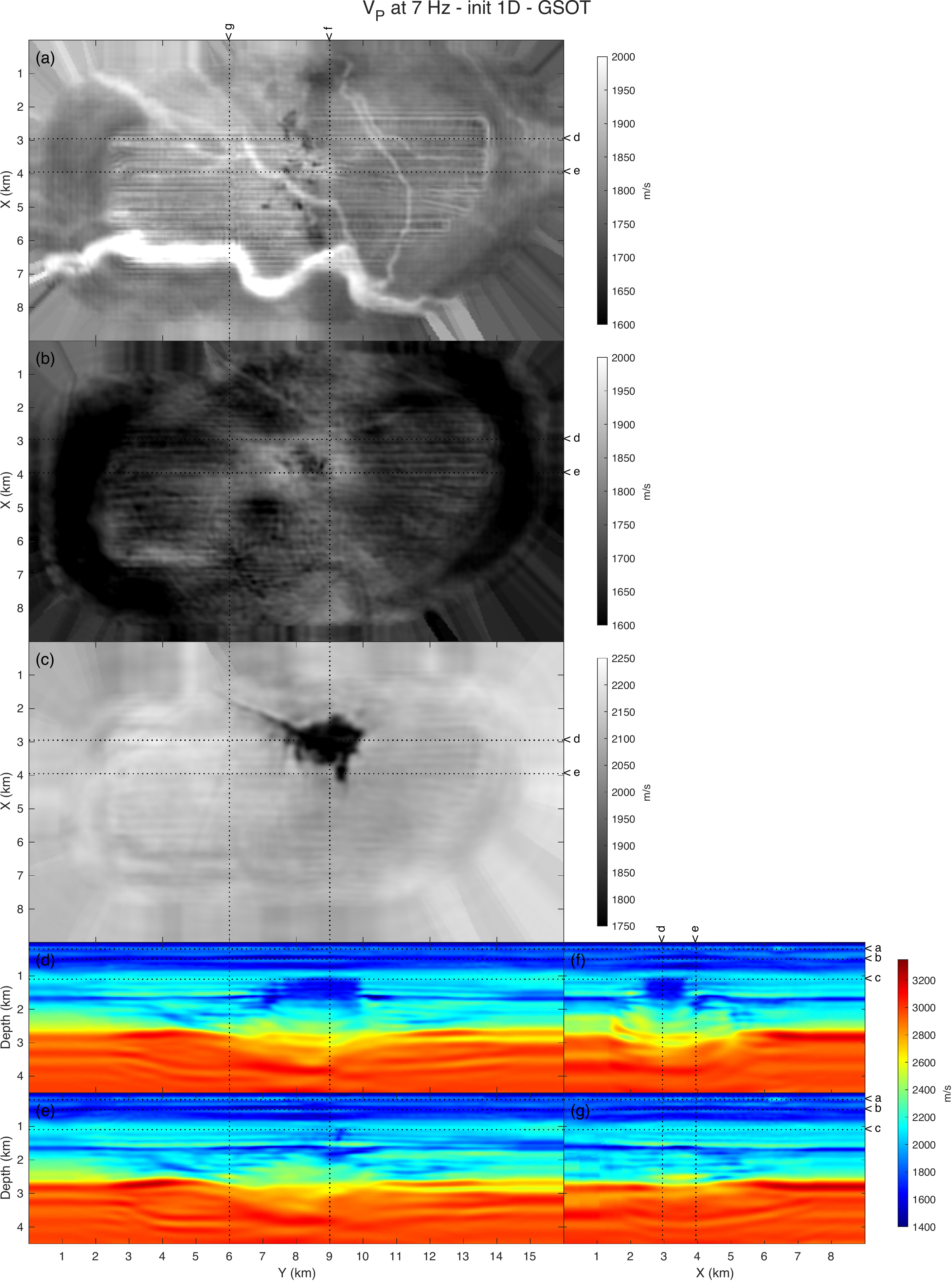}
   \caption{Slices of the $7$~Hz FWI reconstructed velocity using the graph-space OT approach starting from the rough initial model. (a-c) Horizontal slices at (a) $0.2$ km depth, (b) $0.5$ km depth and (c) $1.1$ km depth. (d-e) Inline vertical slices for (d) $x=2.95$~km and (e) $x=3.95$~km. (f-g) Cross-line vertical slices at (f) $y=9$~km and (g) $y=6$~km.}
   \label{vp:VP_init1D_7Hz_GSOT}
\end{figure}

For comparison, the results obtained using the graph-space OT approach starting from the same initial model are presented in Figure \ref{vp:VP_init1D_7Hz_GSOT}. This time the workflow comprises the two frequency bands. One can see that the results obtained, down to $2$ km are similar to the reference results obtained from the accurate initial tomography model. This is particularly encouraging: compared with the least-squares inversion, graph-space OT FWI is able to compensate for the kinematic inaccuracies of the initial model and provides a meaningful velocity reconstruction within the zone of the medium which is sampled both by diving and reflected waves. It is well known that reconstructing the deeper velocity, in a zone sampled exclusively by reflected waves, is a challenge which requires specific treatment (namely treating reflections separately to compute long-wavelength velocity updates from them in the framework of reflection FWI \cite{Xu_2012_IRS,Brossier_2014_VMB,Zhou_2015_FWI,Giuseppe_2020_RGS}). One difference remains: the low velocity anomaly, interpreted as trapped gas, appears slightly deeper than in the reference results (110 m deeper). This is due to an imperfect reconstruction of the shallower part of the medium, which leads to a depth-shifting of this low-velocity anomaly. Note, however, that this corresponds to a $2$ to $3$ grid points difference on a Cartesian grid at $50$ m.\\

\begin{figure}[!ht]
   \centering
   \includegraphics[width=0.9\linewidth]{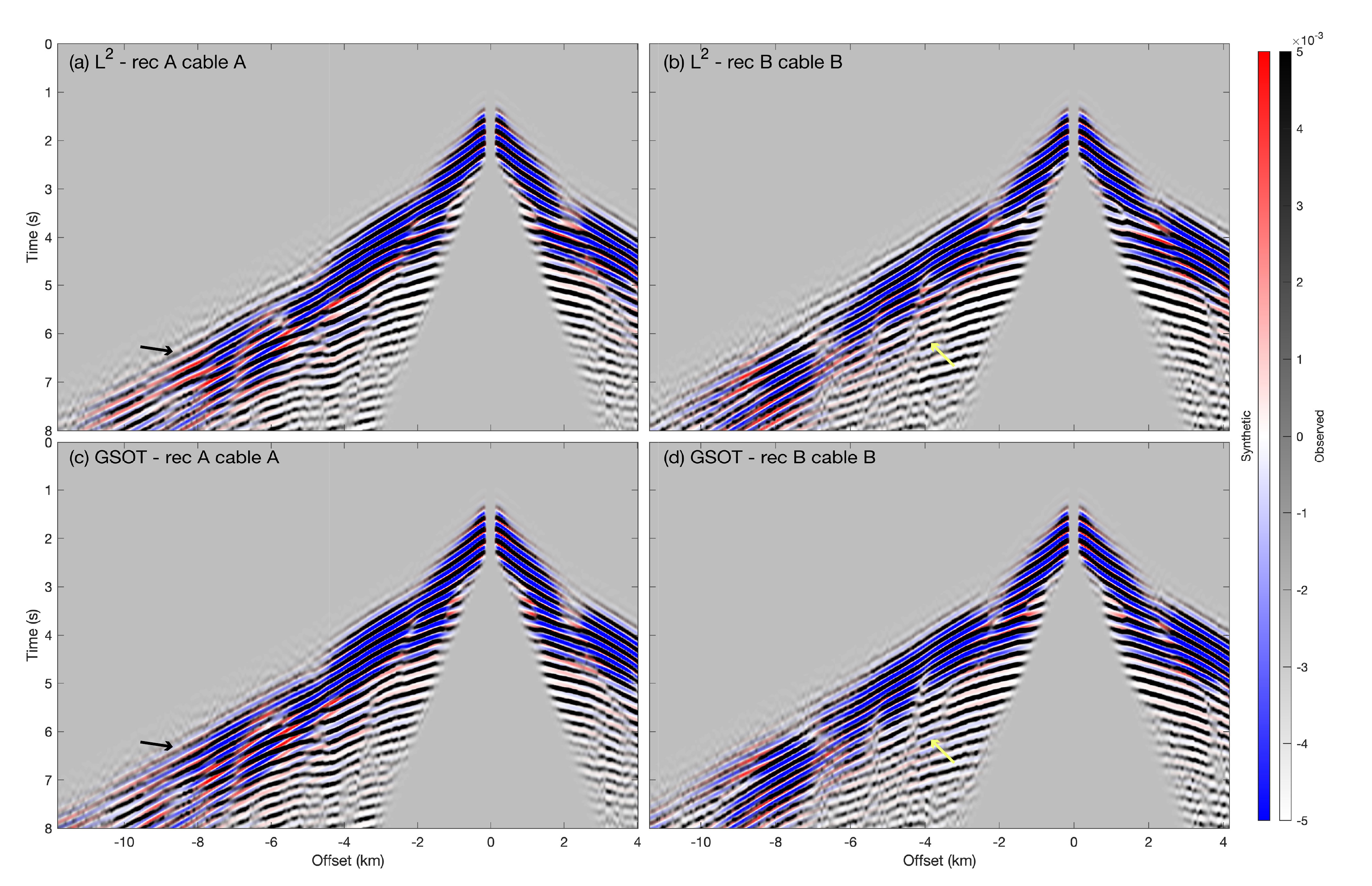}
   \caption{2D common-receiver gathers at $5$~Hz starting from the rough initial model. Synthetic data (blue/white/red color scale) generated into the final velocity model using: (a,b) the leas-squares misfit function, (c,d) the graph-space OT misfit function. (a,c) receiver A along cable A (through the low velocity anomaly). (b,d) receiver B along cable B. Field data are overlapped in grayscale with transparency. Black arrows point to area where graph-space improves the fit to the data.}
   \label{crg:CRG_5Hz_init1D_L2-GS}
\end{figure}

\begin{figure}[!ht]
   \centering
   \includegraphics[width=0.6\linewidth]{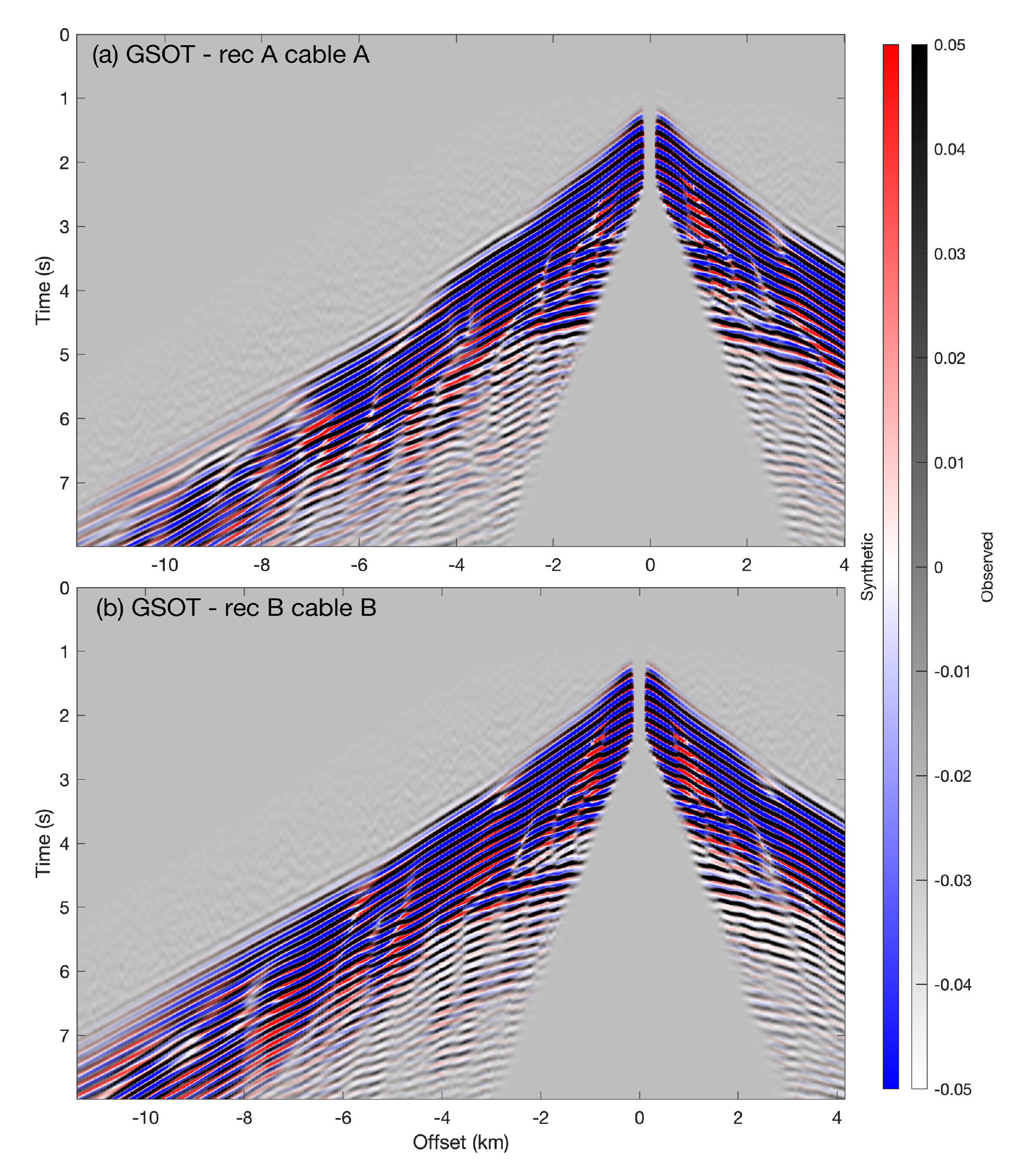}
   \caption{2D common-receiver gathers at $7$~Hz starting from the rough initial model. Synthetic data (blue/white/red color scale)  generated into the final reconstructed velocity using the graph-space OT approach. (a) receiver A along cable A (through the low velocity anomaly). (b) receiver B along cable B. Field data are overlapped in gray-scale with transparency.}
   \label{crg:CRG_7Hz_init1D_GS}
\end{figure}

For comparison, the fit to the data at the first frequency band using least-squares and graph-space OT FWI is presented in Figure \ref{crg:CRG_5Hz_init1D_L2-GS}. One can clearly see the improvement yielded by the graph-space OT strategy over least-squares based FWI. To complete the study, the final data-fit at the second frequency band using the graph-space OT strategy is presented in Figure \ref{crg:CRG_7Hz_init1D_GS}, where it can be seen that the calculated data is in phase with the field data.  

\newpage
\clearpage
\section{Conclusion and perspectives}

The applications of OT distances in the framework of FWI are now well established and have proven their benefits for practical large-scale applications in an industrial context. 
We have reviewed two OT-based methods which are robust when applied to seismic data while improving the convexity of the FWI problem, i.e. alleviating the sensitivity to the initial model and to various conventional workflow steps.\\

The first of these methods, named KR norm-based OT, relies on a specific dual form of the OT distance and has a close connection with the KR norm. Its main benefits in the framework of FWI are its ability to consider the seismic data (or at least data lines) as a whole, accounting for the lateral coherency of the events, to reduce sensitivity to the amplitude information, and to better exploit the low-frequency information in the data. These features, which have been illustrated on the KR norm adjoint source, enhance the general convexity of the FWI problem.
The enhancement of the convexity specifically with respect to time-shifts exists but remains limited.\\

The second method, named graph-space OT, is based on a transformation of each seismic trace into 2D point clouds.
Using such a transform into an OT distance leads to a formalism that allows to greatly improve the convexity with respect to time-shifts. The underlying mechanism produces shifted events in the graph-space adjoint source through a permutation, which has been illustrated.\\

Graph-space and KR FWI thus both have their strengths, which are related to complementary features that  reinforce the kinematic content in the adjoint-source (shifting events for graph-space, and enhancing the amplitudes balancing, low frequencies, and events continuity for KR).\\

The features of graph-space and KR FWI have been illustrated on synthetic tests.
Interestingly, we did not find Marmousi 2 configurations where graph-space FWI outperformed KR FWI or vice versa.
It seems in this case that both graph-space and KR FWI manage to mitigate the non-convexity issues to a similar level, while working very differently on the data.
Such a behavior has also been observed on field data \cite{Kpadonou2021} and deserves further fundamental investigations.\\

Then, 3D field data results were presented.
Several industrial case studies, including land and marine data acquisitions, have shown that KR FWI outperforms least-squares FWI, mitigating non-convexity issues with the specific strengths of KR-based OT.
A marine case study has shown how graph-space FWI outperforms least-squares FWI, also mitigating non-convexity issues with the specific strengths of graph-space OT.\\

A natural perspective regarding the techniques presented here would be to find a way to combine the KR and graph-space approaches to accumulate their respective strengths and bring even more convexity.
A first investigation in this direction has been performed in \cite{Messud2021,Kpadonou2021}, with the proposal of embedding the graph transform into the KR norm.
More investigations are ongoing.\\

Another perspective would be to increase the effective dimensionality of the KR problem (considering a full 3D data representation space instead of a 2D splitting per line)
or of the graph-space problem (considering more than one trace in the graph transform).

\section*{Acknowledgments}

We are grateful to CGG for granting permission to publish this work. We are grateful to PDO, the Ministry of Oil and Gas of the Sultanate of Oman, INEOS, CGG Multi-Client and TGS for granting permission to present the field data results in Section \ref{sec:KR_industrial}.\\

This study was partially funded by the SEISCOPE consortium (\textit{http://seiscope2.osug.fr}), sponsored by AKERBP, CGG, CHEVRON, EQUINOR, EXXON-MOBIL, JGI, SHELL, SINOPEC, SISPROBE and TOTAL. This study was granted access to the HPC resources of the Froggy platform of the CIMENT infrastructure (\textit{https://ciment.ujf-grenoble.fr}), which is supported by the Rh\^one-Alpes region (GRANT CPER07\_13 CIRA), the OSUG@2020 labex (reference ANR10 LABX56) and the Equip@Meso project (reference ANR-10-EQPX-29-01) of the programme Investissements d'Avenir supervised by the Agence Nationale pour la Recherche, and the HPC resources of CINES/IDRIS/TGCC under the allocation 046091 made by GENCI."

\newpage
\clearpage
\begin{appendix}
 
%  \section{Adjoint state technique for gradient computation}
%  \label{app:adjoint}
%  In this section we recall how to derive the adjoint state forma
%  
%  We consider the misfit function
%  \begin{equation}
%  f(m)=\frac{1}{2}\sum_{s=1}^{N_s}F (d_{cal,s}[m]-d_{obs,s}), 
%  \end{equation}

\section{Numerical computation for the Kantorovich-Rubinstein norm}
\label{app:KR}
In this section we assume that the dimension $d$ is set to $3$. Assuming the functions are discretized on a Cartesian mesh with $N$ points $(x_i,y_j,z_k)$, and a spacing $h$ between adjacent point, the discrete problem associated with (\ref{eq:W1dual2}) writes
\begin{equation}
\label{eq:KRdis}
\begin{array}{l}
 \displaystyle
 \max_{\varphi_{ijk}}
\sum_{ijk} \varphi_{ijk} \left(\mu_{ijk} -\nu_{ijk}\right), \;\; s.c. \\
\left\{
 \begin{array}{ll}
  \displaystyle
\forall\; (i,j,k),\; (l,m,n),  &|\varphi_{ijk}-\varphi_{lmn}|< |x_i-x_l|+|y_j-y_m|+|z_k-z_n|, \\
 \displaystyle
 \forall\; (i,j,k), &|\varphi_{ijk}| \leq \lambda.
 \end{array}
\right.
\end{array}
\vspace{1em}
\end{equation}
We use a property of the $\ell_1$ norm on $\mathbb{R}^{d}$ to reduce the number of constraints from $N^2+N$ to $4N$. 
\\
\noindent \textbf{Proposition.}
\\
\textit{
The two following assertions are equivalent
\begin{equation}
\label{eq:prop}
 \begin{array}{ll}
(A1)& 
\displaystyle \forall\; (i,j,k),(l,m,n), \;
\displaystyle
|\varphi_{ijk}-\varphi_{lmn}|< |x_i-x_l|+|y_j-y_m|+|z_k-z_n|, 
\\    
\\
(A2)&
\left\{
\begin{array}{ll}
 \displaystyle \forall\; (i,j,k), \;&\; |\varphi_{i+1,j,k}-\varphi_{ijk}|< |x_{i+1}-x_i|,  \\
 \displaystyle \forall\; (i,j,k), &\; |\varphi_{i,j+1,k}-\varphi_{ijk}|< |y_{j+1}-y_j|,  \\
 \displaystyle \forall\; (i,j,k), &\; |\varphi_{i,j,k+1}-\varphi_{ijk}|< |z_{k+1}-z_k|.
\end{array}
\right.
\end{array}
\end{equation}
}
\vspace{1em}

\noindent \textbf{Proof.}
\\
$(A1)$ obviously implies $(A2)$. To prove the reciprocal implication, consider a pair of points on the mesh denoted by $u$ and $v$, such that 
\begin{equation}
 u=(x_i,y_j,z_k), \;\;
 v=(x_l,y_m,z_n).
\end{equation} 
A sequence of $M \in \mathbb{N}$ points $w_q=(x_{i_q},y_{j_q},z_{k_q}), \; q=1,\dots,M$ can be selected to form a path on the mesh from $u$ to $v$, such that $w_1=u$, $w_M=v$, and $w_q$ are all adjacent on the grid, with monotonically varying coordinates. The key is to see that, for such a sequence of points, the $\ell_1$ norm on $\mathbb{R}^{d}$ ensures that 
\begin{equation}
\label{proof1}
 |v-u| = \sum_{q=1}^{M} |w_{q+1}-w_q|.
\end{equation}
This property of the $\ell_1$ norm is also known as its Manhattan property.

Now, consider a function $\varphi$ satisfying $(A2)$. The triangle inequality yields
\begin{equation}
\label{proof2}
 |\varphi(v)-\varphi(u)| \leq \sum_{q=1}^{M} |\varphi(w_{q+1})-\varphi(w_q)|.
\end{equation}
As the points $w_q$ are adjacent, the local inequalities described by $(A2)$, satisfied by $\varphi$, yield
\begin{equation}
\label{proof3}
 \sum_{q=1}^{M}|\varphi(w_{q+1})-\varphi(w_q)| \leq \sum_{q=1}^{M}|w_{q+1}-w_q|.
\end{equation}
Putting together equations (\ref{proof2}), (\ref{proof3}) and (\ref{proof1}) yields 
\begin{equation}
\label{proof4}
 |\varphi(v)-\varphi(u)| \leq |v-u|,
\end{equation}
or
\begin{equation}
|\varphi_{ijk}-\varphi_{lmn}|< |x_i-x_l|+|y_j-y_m|+|z_k-z_n|,
\end{equation}
which proves the proposition.
\begin{flushright}
$\square$
\end{flushright}
Using the equivalence (\ref{eq:prop}), the problem (\ref{eq:KRdis}) can be rewritten in its equivalent form
\begin{equation}
\label{eq:KRdis2}
\begin{array}{l}
 \displaystyle
 \max_{\varphi_{ijk}}
\sum_{ijk} \varphi_{ijk} \left(\mu_{ijk} -\nu_{ijk}\right), \;\; s.c. 
\\
 \displaystyle
\left\{
\begin{array}{llll}
 \displaystyle \forall\; (i,j,k), \;&\; |\varphi_{,i+1,jk}-\varphi_{ijk}|< |x_{i+1}-x_i|=h_x,  \\
 \displaystyle \forall\; (i,j,k), &\; |\varphi_{i,j+1,k}-\varphi_{ijk}|< |y_{j+1}-y_j|=h_y,  \\
 \displaystyle \forall\; (i,j,k), &\; |\varphi_{i,j,k+1}-\varphi_{ijk}|< |z_{k+1}-z_k|=h_z, \\
 \displaystyle \forall\; (i,j,k), &\; |\varphi_{ijk}|< \lambda. \\  
\end{array}
\right.
\end{array}
\end{equation} 
The problem (\ref{eq:KRdis2}) is equivalent to (\ref{eq:KRdis}) with only $4 N$ constraints, as announced.\\

We solve problem (\ref{eq:KRdis2}) through a proximal splitting algorithm named Alternative Direction Method of Multipliers (ADMM). We first reformulate (\ref{eq:KRdis2}) as the convex non-smooth problem
\begin{equation}
\label{eq:KRconv}
 \max_{\varphi} f_1(\varphi)+f_2(\varphi),
\end{equation}
where 
\begin{equation}
 f_1(\varphi)=\sum_{i,j,k} \varphi_{ijk} \left(\mu_{ijk} -\nu_{ijk}\right), \;\;
f_2=i_{K}\circ A,
\end{equation}
with $K$ the unit hypercube 
\begin{equation}
K=\left\{x \in \mathbb{R}^{4N}, \; |x_i|\leq1, \; i=1,\dots 4N\right\},
\end{equation}
$i_{K}$ the indicator function of $K$
\begin{equation}
 i_{K}(x)=
\left|
\begin{array}{lll}
       0 &\;\;\textrm{if} \;\;& x \in K
\\
       +\infty &\;\;\textrm{if} \;\;& x \notin K,
\end{array}
\right.
\end{equation}
and $A \in \mathbb{M}_{4N,N}(\mathbb{R})$ a rectangular real matrix with $4N$ rows and $N$ columns such that 
\begin{equation}
 A=\left[D_{x}\;\; D_{y}\;\;D_{z}\;\; \frac{1}{\lambda}I_N \right]^{T},
\end{equation}
where $I_N$ is the real identity matrix of size $N$ and $D_x,D_y,D_z$ are the forward finite differences operators
\begin{equation}
\left\{
\begin{array}{c}
\displaystyle
 \left(D_x\varphi\right)_{ijk}=
\displaystyle
\frac{\varphi_{i+1,j,k}-\varphi_{ijk}}{h_x}, 
 \\
\displaystyle
 \left(D_y\varphi\right )_{ijk}=
\displaystyle
\frac{\varphi_{i,j+1,k}-\varphi_{ijk}}{h_y}, 
 \\
\displaystyle
 \left(D_z\varphi\right )_{ijk}=
\displaystyle
\frac{\varphi_{i,j,k+1}-\varphi_{ijk}}{h_z}.
\end{array}
\right.
\end{equation}
The second part of the misfit function $f_2(\varphi)$ represents the constraints of the problem \ref{eq:KRdis2}.\\

The ADMM method can be  described as follows \cite{Combettes_2011_PRO}. 
\begin{algorithm}{}
 \label{algo:sdmm}
%  \SetLine\;
 \caption{ADMM method for the solution of the problem (\ref{eq:KRconv}).}
 $\gamma>0$, $y_1^{0}=0$, $y_2^{0}=0$, $z_1^{0}=0$, $z_2^{0}=0$\;
  \For{$n=0,1,\dots$}{
     $\varphi^n=\left(I_N+A^{T}A\right)^{-1}\left[\left(y_1^n-z_1^n\right)+A^{T}\left(y_2^n-z_2^n\right)\right]$\;
     $y_1^{n+1}=\textrm{prox}_{\gamma f_1} \left(\varphi^n+z_1^n\right)$ \;
     $z_1^{n+1}=z_1^n+\varphi^n-y_1^{n+1}$ \;
     $y_2^{n+1}=\textrm{prox}_{\gamma i_{K}} \left(A\varphi^n+z_2^n\right)$ \;
     $z_2^{n+1}=z_2^n+A\varphi^n-y_2^{n+1}$ \;   
  } 
\end{algorithm} 

Proximal splitting strategies rely on a splitting of the problem in terms of the functions $f_1(\varphi)$ and $f_2(\varphi)$ and the computation of the proximity operators of these two functions (scaled by a positive factor $\gamma$). For the particular case of the function $f_1$ and $i_{K}$, closed-form formulations can be found such that 
\begin{equation}
\label{prox1}
 \textrm{prox}_{\gamma f_1}(\varphi)=\varphi- \gamma(\mu+\nu),
\end{equation}
\begin{equation}
\label{prox2}
 \forall i=1,\dots,4N, \;\; \left(\textrm{prox}_{\gamma i_{K}}(x)\right)_i=
\left|
\begin{array}{cll}
 x_i &\textrm{if}& -1\leq x_i\leq 1
 \\
 1 &\textrm{if}& x_i> 1
 \\
 -1&\textrm{if}& x_i<- 1.
\end{array}
\right.
\end{equation}
The closed-form formulations (\ref{prox1}) and (\ref{prox2}) are inexpensive to compute with an overall complexity in $O(N)$ operations.\\

However, the ADMM algorithm requires the solution of a linear system involving the matrix $I+A^{T}A$, which is the most time-consuming part of the algorithm. We have
\begin{equation}
 A^{T}A=\Delta + \frac{1}{\lambda^2}I_N, \;\; \Delta= D_x^{T}D_x+D_y^{T}D_y+D_z^{T}D_z. 
\end{equation}
In \cite{Metivier_2016_OTI} we prove that the matrix $\Delta$ actually corresponds to the second-order finite differences discretization of the 3D Laplacian operator defined on $\Omega$ with homogeneous Neumann boundary conditions. We redirect the reader to this study for a formal proof.\\

The linear system which has to be solved at each iteration of the ADMM algorithm thus corresponds to a second-order finite-differences discretization of the  Poisson's problem
\begin{equation}
\label{poisson}
 -\left(\Delta +\left(1+\frac{1}{\lambda^2}\right)I_N \right) \varphi^n =f^n,
\end{equation}
where $\Delta$ is a Laplacian operator with homogeneous Neumann boundary conditions and $f^n=-\left(y_1^n-z_1^n\right)-A^{T}\left(y_2^n-z_2^n\right)$. The best numerical strategies for the solution of such problems appears to rely either on the Fast Fourier Transform algorithm with $O(N\log N)$ complexity \cite{Swarztrauber_1974_FIS} or multigrid solvers with $O(N)$ complexity \cite{Brandt_1977_MUL}.\\

The combination of the reduction of the number of constraints using the property of the $\ell_1$ distance and the observation that the matrix appearing in the ADMM strategy actually corresponds to the discretization of the Poisson's equation offers the possibility to design an efficient numerical method to compute the KR norm for large scale problems.

\end{appendix}


\newcommand{\SortNoop}[1]{}
\begin{thebibliography}{100}

\bibitem{Aghamiry_2019_IWR}
H.~Aghamiry, A.~Gholami, and S.~Operto.
\newblock Improving full-waveform inversion by wavefield reconstruction with
  alternating direction method of multipliers.
\newblock {\em Geophysics}, 84(1):R139--R162, 2019.

\bibitem{Akgul_1993_GPP}
Mustafa Akg\"ul.
\newblock A genuinely polynomial primal simplex algorithm for the assignment
  problem.
\newblock {\em Discrete Applied Mathematics}, 45(2):93--115, 1993.

\bibitem{Aki_1980_QST}
K.~Aki and P.~Richards.
\newblock {\em Quantitative Seismology: Theory and Methods}.
\newblock W. H. Freeman \& Co, San Francisco, 1980.

\bibitem{Ambrosio_2003}
Luigi Ambrosio.
\newblock Lecture notes on optimal transport problems.
\newblock In {\em Mathematical Aspects of Evolving Interfaces}, volume 1812 of
  {\em Lecture Notes in Mathematics}, pages 1--52. Springer Berlin Heidelberg,
  2003.

\bibitem{Ambrosio_2011_GFS}
Luigi Ambrosio, Edoardo Mainini, and Sylvia Serfaty.
\newblock Gradient flow of the {C}hapman {R}ubinstein {S}chatzman model for
  signed vortices.
\newblock {\em Annales de l'Institut Henri Poincar\'e (C) Non Linear Analysis},
  28(2):217--246, 2011.

\bibitem{Balinski_1985_SMA}
M.~L. Balinski.
\newblock Signature methods for the assignment problem.
\newblock {\em Operations Research}, 33(3):527--536, 1985.

\bibitem{Barkved_2003_LoFS}
O.I. Barkved, A.G. B{\ae}rheim, D.J. Howe, J.H. Kommedal, and G.~Nicol.
\newblock Life of {F}ield {S}eismic {I}mplementation - {A}nother ``first at
  valhal''.
\newblock In {\em 65$^{th}$ EAGE Workshop, Stavanger}, 2003.

\bibitem{Bertsekas_1989_AAT}
D.~P. Bertsekas and D.A. Castanon.
\newblock The auction algorithm for the transportation problem.
\newblock {\em Annals of Operations Research}, 20(1):67--96, 1989.

\bibitem{Bertsekas_1998_NOC}
Dimitri~P. Bertsekas.
\newblock {\em Network Optimization: Continuous and Discrete Models}.
\newblock Athena Scientific, 1998.

\bibitem{Bleistein_1987_IRE}
N.~Bleistein.
\newblock On the imaging of reflectors in the {E}arth.
\newblock {\em Geophysics}, 52(7):931--942, 1987.

\bibitem{Bogachev_2007_MET}
V.~I. Bogachev.
\newblock {\em Measure Theory}.
\newblock Number vol.~I,II in Measure Theory. Springer Berlin Heidelberg, 2007.

\bibitem{Bozdag_2016_GAT}
Ebru Bozda{\u{g}}, Daniel Peter, Matthieu Lefebvre, Dimitri Komatitsch, Jeroen
  Tromp, Judith Hill, Norbert Podhorszki, and David Pugmire.
\newblock Global adjoint tomography: first-generation model.
\newblock {\em Geophysical Journal International}, 207(3):1739--1766, 2016.

\bibitem{Bozdag_2011_MFF}
E.~Bozda\u{g}, J.~Trampert, and J.~Tromp.
\newblock Misfit functions for full waveform inversion based on instantaneous
  phase and envelope measurements.
\newblock {\em Geophysical Journal International}, 185(2):845--870, 2011.

\bibitem{Brandt_1977_MUL}
A.~Brandt.
\newblock Multi-level adaptive solutions to boundary-value problems.
\newblock {\em Mathematics of Computation}, 31:333--390, 1977.

\bibitem{Bretaudeau_2013_EFW}
F.~Bretaudeau, R.~Brossier, D.~Leparoux, O.~Abraham, and J.~Virieux.
\newblock {2D elastic full waveform imaging of the near surface: Application to
  synthetic and a physical modelling data sets}.
\newblock {\em Near Surface Geophysics}, 11:307--316, 2013.

\bibitem{Brossier_2009_SIC}
R.~Brossier, S.~Operto, and J.~Virieux.
\newblock Seismic imaging of complex onshore structures by {2D} elastic
  frequency-domain full-waveform inversion.
\newblock {\em Geophysics}, 74(6):WCC105--WCC118, 2009.

\bibitem{Brossier_2014_VMB}
R.~Brossier, S.~Operto, and J.~Virieux.
\newblock Velocity model building from seismic reflection data by full waveform
  inversion.
\newblock {\em Geophysical Prospecting}, 63:354--367, 2015.

\bibitem{Bunks_1995_MSW}
C.~Bunks, F.~M. Salek, S.~Zaleski, and G.~Chavent.
\newblock Multiscale seismic waveform inversion.
\newblock {\em Geophysics}, 60(5):1457--1473, 1995.

\bibitem{Burkard_2012_APR}
R.~Burkard, M.~Dell'Amico, and S.~Martello.
\newblock {\em Assignment {P}roblems}.
\newblock Society for Industrial and Applied Mathematics, 2012.

\bibitem{Carotti2020}
D.~Carotti, O.~Hermant, S.~Masclet, M.~Reinier, J.~Messud, A.~Sedova, and
  G.~Lambar\'e.
\newblock Optimal transport full-waveform inversion - {Applications}.
\newblock {\em 82nd EAGE Conference and Exhibition, Expanded Abstracts}, Th
  Dome1 17, 2020.

\bibitem{Claerbout_1985_IEI}
J.F. Claerbout.
\newblock {\em Imaging the {E}arth's interior}.
\newblock Blackwell Scientific Publication, 1985.

\bibitem{Combettes_2011_PRO}
P.~L. Combettes and J-C. Pesquet.
\newblock Proximal splitting methods in signal processing.
\newblock In H.~H. Bauschke, R.~S. Burachik, P.~L. Combettes, V.~Elser, D.~R.
  Luke, and H.~Wolkowicz, editors, {\em Fixed-Point Algorithms for Inverse
  Problems in Science and Engineering}, volume~49 of {\em Springer Optimization
  and Its Applications}, pages 185--212. Springer New York, 2011.

\bibitem{Delon_2006_MVS}
J.~Delon.
\newblock Movie and video scale-time equalization application to flicker
  reduction.
\newblock {\em IEEE Transactions on Image Processing}, 15(1):241--248, Jan
  2006.

\bibitem{Dominitz_2010_TMO}
Ayelet Dominitz and Allen Tannenbaum.
\newblock Texture mapping via optimal mass transport.
\newblock {\em IEEE Transactions on Visualization and Computer Graphics},
  16(3):419--433, May 2010.

\bibitem{Engquist_2014_WAS}
B.~Engquist and B.~D. Froese.
\newblock Application of the {W}asserstein metric to seismic signals.
\newblock {\em Communications in Mathematical Science}, 12(5):979--988, 2014.

\bibitem{Engquist_2016_WAS}
B.~Engquist, B.~D. Froese, and Y.~Yang.
\newblock Optimal transport for seismic full waveform inversion.
\newblock {\em Communications in Mathematical Sciences}, 14(8):2309--2330,
  2016.

\bibitem{Fichtner_2008_TBC}
A.~Fichtner, B.~L.~N. Kennett, H.~Igel, and H.~P. Bunge.
\newblock Theoretical background for continental- and global-scale
  full-waveform inversion in the time-frequency domain.
\newblock {\em Geophysical Journal International}, 175:665--685, 2008.

\bibitem{Fichtner_2010_FWT}
A.~Fichtner, B.~L.~N. Kennett, H.~Igel, and H.~P. Bunge.
\newblock { Full waveform tomography for radially anisotropic structure: New
  insights into present and past states of the Australasian upper mantle}.
\newblock {\em Earth and Planetary Science Lettters}, 290(3-4):270--280, 2010.

\bibitem{Gauthier_1986_TDN}
O.~Gauthier, J.~Virieux, and A.~Tarantola.
\newblock Two-dimensional nonlinear inversion of seismic waveforms: numerical
  results.
\newblock {\em Geophysics}, 51(7):1387--1403, July 1986.

\bibitem{Groos_2014_RAF}
L.~Groos, M.~Sch\"afer, T.~Forbriger, and T.~Bohlen.
\newblock The role of attenuation in {2D} full-waveform inversion of
  shallow-seismic body and {R}ayleigh waves.
\newblock {\em Geophysics}, 79(6):R247--R261, 2014.

\bibitem{Hermant2020}
O.~Hermant, A.~Aziz, S.~Warzocha, and M.~Al~Jahdhami.
\newblock Imaging complex fault structures on-shore {Oman} using optimal
  transport full-waveform inversion.
\newblock {\em 82nd EAGE Conference and Exhibition, Expanded Abstracts}, We
  Dome1 19, 2020.

\bibitem{Hermant2019}
O.~Hermant, A.~Sedova, G.~Royle, M.~Retailleau, J.~Messud, G.~Lambar\'e,
  S.~Al~Abri, and M.~Al~Jahdhami.
\newblock Broadband {FAZ} land data: an opportunity for {FWI}.
\newblock {\em 81st EAGE Conference and Exhibition, Workshop}, WS08 11, 2019.

\bibitem{Huang_2019_WIS}
Guanghui Huang, Rami Nammour, William~W. Symes, and Mohamed Dolliazal.
\newblock {\em Waveform inversion via source extension}, pages 4761--4766.
\newblock 2019.

\bibitem{Irnaka_2019_T3D}
T.M. Irnaka, R.~Brossier, L.~M\'etivier, T.~Bohlen, and Y.~Pan.
\newblock Towards 3d 9c elastic full waveform inversion of shallow seismic
  wavefields - case study ettlingen line.
\newblock In {\em Expanded Abstracts, 81$^{th}$ Annual {EAGE} Conference \&
  Exhibition, London}, page We P01 04. EAGE, 2019.

\bibitem{Jannane_1989_WES}
M.~Jannane, W.~Beydoun, E.~Crase, D.~Cao, Z.~Koren, E.~Landa, M.~Mendes,
  A.~Pica, M.~Noble, G.~Roeth, S.~Singh, R.~Snieder, A.~Tarantola, and
  D.~Trezeguet.
\newblock Wavelengths of {Earth} structures that can be resolved from seismic
  reflection data.
\newblock {\em Geophysics}, 54(7):906--910, 1989.

\bibitem{Kamath_2020_MFW}
N.~Kamath, R.~Brossier, L.~M\'etivier, A.~Pladys, and P.~Yang.
\newblock Multiparameter full-waveform inversion of {3D} ocean-bottom cable
  data from the {V}alhall field.
\newblock {\em Geophysics}, 86(1):B15--B35, 2021.

\bibitem{Kantorovich_1942_TOM}
L.~Kantorovich.
\newblock On the transfer of masses.
\newblock {\em Dokl. Acad. Nauk. USSR}, 37:7--8, 1942.

\bibitem{Kpadonou2021}
F.~Kpadonou, J.~Messud, An. Sedova, and M.~Reinier.
\newblock Optimal transport {FWI} with graph transform: Analysis and proposal
  of a partial shift strategy.
\newblock {\em 83rd EAGE Conference and Exhibition}, 2021.

\bibitem{Kunh_1955_HMA}
H.~W. Kuhn.
\newblock The {H}ungarian method for the assignment problem.
\newblock {\em Naval Research Logistics Quarterly}, 2(1-2):83--97, 1955.

\bibitem{Lailly_1983_SIP}
P.~Lailly.
\newblock The seismic inverse problem as a sequence of before stack migrations.
\newblock In Robinson Bednar and Weglein, editors, {\em Conference on {I}nverse
  {S}cattering, Theory and application, Society for Industrial and Applied
  Mathematics, Philadelphia}, pages 206--220, 1983.

\bibitem{Lambare_2008_STE}
G.~Lambar\'e.
\newblock Stereotomography.
\newblock {\em Geophysics}, 73(5):VE25--VE34, 2008.

\bibitem{Lellmann_2014_KRU}
J.~Lellmann, D.A. Lorenz, C.~{Sch\" onlieb}, and T.~Valkonen.
\newblock Imaging with {K}antorovich--{R}ubinstein discrepancy.
\newblock {\em SIAM Journal on Imaging Sciences}, 7(4):2833--2859, 2014.

\bibitem{Luo_2011_DBO}
Simon Luo and Paul Sava.
\newblock A deconvolution-based objective function for wave-equation inversion.
\newblock {\em SEG Technical Program Expanded Abstracts}, 30(1):2788--2792,
  2011.

\bibitem{Luo_1991_WET}
Y.~Luo and G.~T. Schuster.
\newblock Wave-equation traveltime inversion.
\newblock {\em Geophysics}, 56(5):645--653, 1991.

\bibitem{Mainini_2012_DTC}
E.~Mainini.
\newblock A description of transport cost for signed measures.
\newblock {\em Journal of Mathematical Sciences}, 181(6):837--855, 2012.

\bibitem{Martin_2006_M2E}
G.~S. Martin, R.~Wiley, and K.~J. Marfurt.
\newblock Marmousi2: An elastic upgrade for {M}armousi.
\newblock {\em The Leading Edge}, 25(2):156--166, 2006.

\bibitem{Messud2021}
J.~Messud, R.~Poncet, and G.~Lambar\'e.
\newblock Optimal transport in full-waveform inversion: Analysis and practice
  of the multidimensional kantorovich-rubinstein norm.
\newblock {\em Inverse Problems}, 37(065012):1--42, 2021.

\bibitem{Messud2019}
J.~Messud and A.~Sedova.
\newblock Multidimensional optimal transport for {3D FWI}: Demonstration on
  field data.
\newblock {\em 81st EAGE Conference and Exhibition, Expanded Abstracts}, Tu R08
  02, 2019.

\bibitem{Metivier_2018_OTD}
L.~M{\'e}tivier, A.~Allain, R.~Brossier, Q.~M{\'e}rigot, E.~Oudet, and
  J.~Virieux.
\newblock {\em On the Use of Optimal Transport Distances for a PDE-Constrained
  Optimization Problem in Seismic Imaging}, pages 377--397.
\newblock Springer New York, New York, NY, 2018.

\bibitem{Metivier_2018_OTM}
L.~M\'etivier, A.~Allain, R.~Brossier, Q.~M\'erigot, E.~Oudet, and J.~Virieux.
\newblock Optimal transport for mitigating cycle skipping in full waveform
  inversion: a graph space transform approach.
\newblock {\em Geophysics}, 83(5):R515--R540, 2018.

\bibitem{Metivier_2021_NGS}
L.~M\'etivier and R.~Brossier.
\newblock New insights on the graph space optimal transport distance for full
  waveform inversion.
\newblock In {\em SEG Technical Program Expanded Abstracts 2021}, 2021.

\bibitem{Metivier_2019_GOT}
L.~M\'etivier, R.~Brossier, Q.~M\'erigot, and E.~Oudet.
\newblock A graph space optimal transport distance as a generalization of
  ${L}^p$ distances: application to a seismic imaging inverse problem.
\newblock {\em Inverse Problems}, 35(8):085001, 2019.

\bibitem{Metivier_2016_TLE}
L.~M\'etivier, R.~Brossier, Q.~M\'erigot, E.~Oudet, and J.~Virieux.
\newblock Increasing the robustness and applicability of full waveform
  inversion: an optimal transport distance strategy.
\newblock {\em The Leading Edge}, 35(12):1060--1067, 2016.

\bibitem{Metivier_2016_TOF}
L.~M\'etivier, R.~Brossier, Q.~M\'erigot, E.~Oudet, and J.~Virieux.
\newblock Measuring the misfit between seismograms using an optimal transport
  distance: {A}pplication to full waveform inversion.
\newblock {\em Geophysical Journal International}, 205:345--377, 2016.

\bibitem{Metivier_2016_OTI}
L.~M\'etivier, R.~Brossier, Q.~M\'erigot, E.~Oudet, and J.~Virieux.
\newblock An optimal transport approach for seismic tomography: Application to
  {3D} full waveform inversion.
\newblock {\em Inverse Problems}, 32(11):115008, 2016.

\bibitem{Metivier_2021_REL}
Ludovic M\'etivier and Romain Brossier.
\newblock Receiver-extension strategy for time-domain full-waveform inversion
  using a relocalization approach.
\newblock {\em Geophysics}, 87(1):R13--R33, 2022.

\bibitem{Monge_1781_MSL}
Gaspard Monge.
\newblock M{\'e}moire sur la th{\'e}orie des d{\'e}blais et des remblais.
\newblock {\em Histoire de l'Acad{\'e}mie Royale des Sciences de Paris}, 1781.

\bibitem{Nocedal_1980_UQN}
J.~Nocedal.
\newblock {Updating Quasi-{N}ewton Matrices With Limited Storage}.
\newblock {\em Mathematics of Computation}, 35(151):773--782, 1980.

\bibitem{Nocedal_2006_NO}
J.~Nocedal and S.~J. Wright.
\newblock {\em Numerical Optimization}.
\newblock Springer, 2nd edition, 2006.

\bibitem{Nolet_2008_BST}
G.~Nolet.
\newblock {\em {A Breviary of Seismic Tomography}}.
\newblock Cambridge University Press, Cambridge, UK, 2008.

\bibitem{Operto_2013_TLE}
S.~Operto, R.~Brossier, Y.~Gholami, L.~M\'etivier, V.~Prieux, A.~Ribodetti, and
  J.~Virieux.
\newblock A guided tour of multiparameter full waveform inversion for
  multicomponent data: from theory to practice.
\newblock {\em The Leading Edge}, Special section Full Waveform
  Inversion(September):1040--1054, 2013.

\bibitem{Operto_2015_ETF}
S.~Operto, A.~Miniussi, R.~Brossier, L.~Combe, L.~M\'etivier, V.~Monteiller,
  A.~Ribodetti, and J.~Virieux.
\newblock Efficient {3-D} frequency-domain mono-parameter full-waveform
  inversion of ocean-bottom cable data: application to {V}alhall in the
  visco-acoustic vertical transverse isotropic approximation.
\newblock {\em Geophysical Journal International}, 202(2):1362--1391, 2015.

\bibitem{Pitie_2007_ACG}
François Piti\'e, Anil~C. Kokaram, and Rozenn Dahyot.
\newblock Automated colour grading using colour distribution transfer.
\newblock {\em Computer Vision and Image Understanding}, 107(1):123 -- 137,
  2007.
\newblock Special issue on color image processing.

\bibitem{Pladys_2021_OCM}
A.~Pladys, R.~Brossier, Y.~Li, and L.~M\'etivier.
\newblock On cycle-skipping and misfit function modification for full-wave
  inversion: {C}omparison of five recent approaches.
\newblock {\em Geophysics}, 86(4):R563--R587, 2021.

\bibitem{Pladys_2021_OTV}
Arnaud Pladys, Romain Brossier, Nishant Kamath, and Ludovic M\'etivier.
\newblock Robust {FWI} with graph space optimal transport: application to {3D}
  {OBC} {V}alhall data.
\newblock {\em Geophysics}, 87(3):1--76, 2022.

\bibitem{Plessix_2006_RAS}
R.~E. Plessix.
\newblock A review of the adjoint-state method for computing the gradient of a
  functional with geophysical applications.
\newblock {\em Geophysical Journal International}, 167(2):495--503, 2006.

\bibitem{Plessix_2010_FWI}
R.~E. Plessix and C.~Perkins.
\newblock Full waveform inversion of a deep water ocean bottom seismometer
  dataset.
\newblock {\em First Break}, 28:71--78, 2010.

\bibitem{Poncet2018}
R.~Poncet, J.~Messud, M.~Bader, G.~Lambar\'e, G.~Viguier, and C.~Hidalgo.
\newblock {FWI} with optimal transport: {A} {3D} implementation and an
  application on a field dataset.
\newblock {\em 80th EAGE Conference and Exhibition, Expanded Abstracts}, We A12
  02, 2018.

\bibitem{Pratelli_2007_MKP}
Aldo Pratelli.
\newblock On the equality between {M}onge's infimum and {K}antorovich's minimum
  in optimal mass transportation.
\newblock {\em Annales de l'Institut Henri Poincare (B) Probability and
  Statistics}, 43(1):1 -- 13, 2007.

\bibitem{Prieux_2011_FAI}
V.~Prieux, R.~Brossier, Y.~Gholami, S.~Operto, J.~Virieux, O.I. Barkved, and
  J.H. Kommedal.
\newblock On the footprint of anisotropy on isotropic full waveform inversion:
  the {V}alhall case study.
\newblock {\em Geophysical Journal International}, 187:1495--1515, 2011.

\bibitem{Prieux_2013_MFWa}
V.~Prieux, R.~Brossier, S.~Operto, and J.~Virieux.
\newblock Multiparameter full waveform inversion of multicomponent {OBC} data
  from {V}alhall. {P}art 1: imaging compressional wavespeed, density and
  attenuation.
\newblock {\em Geophysical Journal International}, 194(3):1640--1664, 2013.

\bibitem{Giuseppe_2020_RGS}
Giuseppe Provenzano, Romain Brossier, Ludovic M´etivier, and Yubing Li.
\newblock {\em Joint FWI of diving and reflected waves using a graph space
  optimal transport distance: Synthetic tests on limited-offset surface seismic
  data}, pages 780--784.
\newblock 2020.

\bibitem{Qiu_2017_EOT}
Lingyun Qiu, Jaime Ramos-Mart\`inez, Alejandro Valenciano, Yunan Yang, and
  Bj{\"o}rn Engquist.
\newblock Full-waveform inversion with an exponentially encoded
  optimal-transport norm.
\newblock In {\em SEG Technical Program Expanded Abstracts 2017}, pages
  1286--1290, 2017.

\bibitem{Rabin_2010_GSR}
Julien Rabin, Gabriel Peyr{\'e}, and Laurent~D. Cohen.
\newblock Geodesic {S}hape {R}etrieval via {O}ptimal {M}ass {T}ransport.
\newblock In Kostas Daniilidis, Petros Maragos, and Nikos Paragios, editors,
  {\em Computer Vision -- ECCV 2010}, pages 771--784, Berlin, Heidelberg, 2010.
  Springer Berlin Heidelberg.

\bibitem{Rabin_2012_WBA}
Julien Rabin, Gabriel Peyr{\'e}, Julie Delon, and Marc Bernot.
\newblock {W}asserstein {B}arycenter and {I}ts {A}pplication to {T}exture
  {M}ixing.
\newblock In Alfred~M. Bruckstein, Bart~M. ter Haar~Romeny, Alexander~M.
  Bronstein, and Michael~M. Bronstein, editors, {\em Scale Space and
  Variational Methods in Computer Vision}, pages 435--446, Berlin, Heidelberg,
  2012. Springer Berlin Heidelberg.

\bibitem{Rubner_2000_EMD}
Yossi Rubner, Carlo Tomasi, and Leonidas~J. Guibas.
\newblock The {E}arth mover's distance as a metric for image retrieval.
\newblock {\em International Journal of Computer Vision}, 40(2):99--121, Nov
  2000.

\bibitem{Santambrogio_2015_OAM}
F.~Santambrogio.
\newblock {\em Optimal Transport for Applied Mathematicians: Calculus of
  Variations, PDEs, and Modeling}.
\newblock Progress in Nonlinear Differential Equations and Their Applications.
  Springer International Publishing, 2015.

\bibitem{Schafer_2013_TFW}
M.~Sch\"afer, L.~Groos, T.~Forbriger, and T.~Bohlen.
\newblock {2D} full waveform inversion of recorded shallow seismic {R}ayleigh
  waves on a significantly {2D} structure.
\newblock In {\em Proceedings of 19th European Meeting of Environmental and
  Engineering Geophysics, Expanded Abstracts, Bochum, Germany}, 2013.

\bibitem{Sedova2019}
A.~Sedova, J.~Messud, H.~Prigent, S.~Masclet, G.~Royle, and G.~Lambar\'e.
\newblock Acoustic land full-waveform inversion on a broadband land dataset:
  {The} impact of optimal transport.
\newblock {\em 81st EAGE Conference and Exhibition, Expanded Abstracts}, Th R08
  07, 2019.

\bibitem{Shipp_2002_TDF}
R.~M. Shipp and S.~C. Singh.
\newblock Two-dimensional full wavefield inversion of wide-aperture marine
  seismic streamer data.
\newblock {\em Geophysical Journal International}, 151:325--344, 2002.

\bibitem{Sirgue_2010_FWI}
L.~Sirgue, O.~I. Barkved, J.~Dellinger, J.~Etgen, U.~Albertin, and J.~H.
  Kommedal.
\newblock Full waveform inversion: the next leap forward in imaging at
  {V}alhall.
\newblock {\em First Break}, 28:65--70, 2010.

\bibitem{Stopin_2014_MWI}
A.~Stopin, R.-E. Plessix, and S.~{Al Abri}.
\newblock Multiparameter waveform inversion of a large wide-azimuth
  low-frequency land data set in {O}man.
\newblock {\em Geophysics}, 79(3):WA69--WA77, 2014.

\bibitem{Swarztrauber_1974_FIS}
P.~N. Swarztrauber.
\newblock A {D}irect {M}ethod for the {D}iscrete {S}olution of {S}eparable
  {E}lliptic {E}quations.
\newblock {\em SIAM Journal on Numerical Analysis}, 11(6):1136--1150, 1974.

\bibitem{Symes_2008_MVA}
W.~W. Symes.
\newblock Migration velocity analysis and waveform inversion.
\newblock {\em Geophysical Prospecting}, 56:765--790, 2008.

\bibitem{Tape_2010_STS}
C.~Tape, Q.~Liu, A.~Maggi, and J.~Tromp.
\newblock Seismic tomography of the southern {C}alifornia crust based on
  spectral-element and adjoint methods.
\newblock {\em Geophysical Journal International}, 180:433--462, 2010.

\bibitem{Tarantola_1984_ISR}
A.~Tarantola.
\newblock Inversion of seismic reflection data in the acoustic approximation.
\newblock {\em Geophysics}, 49(8):1259--1266, 1984.

\bibitem{vanLeeuwen_2016_PMP}
T.~{van Leeuwen} and F.~Herrmann.
\newblock A penalty method for {PDE}-constrained optimization in inverse
  problems.
\newblock {\em Inverse Problems}, 32(1):1--26, 2016.

\bibitem{VanLeeuwen_2013_MLM}
T.~{van Leeuwen} and F.~J. Herrmann.
\newblock Mitigating local minima in full-waveform inversion by expanding the
  search space.
\newblock {\em Geophysical Journal International}, 195(1):661--667, 2013.

\bibitem{VanLeeuwen_2010_CMC}
T.~{van Leeuwen} and W.~A. Mulder.
\newblock A correlation-based misfit criterion for wave-equation traveltime
  tomography.
\newblock {\em Geophysical Journal International}, 182(3):1383--1394, 2010.

\bibitem{Villani_2003_AMS}
C.~Villani.
\newblock {\em Topics in optimal transportation}.
\newblock Graduate Studies In Mathematics, {Vol}. 50, {AMS}, 2003.

\bibitem{Villani_2008_OTO}
C.~Villani.
\newblock {\em Optimal transport: old and new}.
\newblock Grundlehren der mathematischen Wissenschaften. Springer, Berlin,
  2008.

\bibitem{Virieux_2017_FWI}
J.~Virieux, A.~Asnaashari, R.~Brossier, L.~M\'etivier, A.~Ribodetti, and
  W.~Zhou.
\newblock An introduction to {F}ull {W}aveform {I}nversion.
\newblock In V.~Grechka and K.~Wapenaar, editors, {\em Encyclopedia of
  Exploration Geophysics}, pages R1--1--R1--40. Society of Exploration
  Geophysics, 2017.

\bibitem{Wang_2009_RSW}
Y.~Wang and Y.~Rao.
\newblock Reflection seismic waveform tomography.
\newblock {\em Journal of Geophysical Research}, 114(B3):1978--2012, 2009.

\bibitem{Warner_2016_AWI}
M.~Warner and L.~Guasch.
\newblock Adaptive waveform inversion: Theory.
\newblock {\em Geophysics}, 81(6):R429--R445, 2016.

\bibitem{Wu_2014_SEI}
Ru-Shan Wu, Jingrui Luo, and Bangyu Wu.
\newblock Seismic envelope inversion and modulation signal model.
\newblock {\em Geophysics}, 79(3):WA13--WA24, 2014.

\bibitem{Xu_2012_IRS}
S.~Xu, D.~Wang, F.~Chen, G.~Lambar\'e, and Y.~Zhang.
\newblock Inversion on reflected seismic wave.
\newblock {\em SEG Technical Program Expanded Abstracts 2012}, pages 1--7,
  2012.

\bibitem{Yang_2018_TRN}
Pengliang Yang, Romain Brossier, Ludovic M\'etivier, Jean Virieux, and Wei
  Zhou.
\newblock A {T}ime-{D}omain {P}reconditioned {T}runcated {N}ewton {A}pproach to
  {M}ultiparameter {V}isco-acoustic {F}ull {W}aveform {I}nversion.
\newblock {\em SIAM Journal on Scientific Computing}, 40(4):B1101--B1130, 2018.

\bibitem{Yang_2018_WAS}
Yunan Yang and Bj\"{o}rn Engquist.
\newblock Analysis of optimal transport and related misfit functions in
  full-waveform inversion.
\newblock {\em GEOPHYSICS}, 83(1):A7--A12, 2018.

\bibitem{Yang_2018_MRB}
Yunan Yang and Bj\"{o}rn Engquist.
\newblock {\em Model recovery below reflectors by optimal-transport FWI}, pages
  1178--1182.
\newblock 2018.

\bibitem{Yang_2019_SEG}
Yunan Yang and Bj\"{o}rn Engquist.
\newblock {\em Improving optimal transport based FWI through data
  normalization}, pages 1315--1319.
\newblock 2019.

\bibitem{Yang_2017_AOT}
Yunan Yang, Bj\"{o}rn Engquist, Junzhe Sun, and Brittany~F. Hamfeldt.
\newblock Application of optimal transport and the quadratic {W}asserstein
  metric to full-waveform inversion.
\newblock {\em Geophysics}, 83(1):R43--R62, 2018.

\bibitem{Zhou_2015_FWI}
W.~Zhou, R.~Brossier, S.~Operto, and J.~Virieux.
\newblock Full waveform inversion of diving \& reflected waves for velocity
  model building with impedance inversion based on scale separation.
\newblock {\em Geophysical Journal International}, 202(3):1535--1554, 2015.

\end{thebibliography}


\newpage
\clearpage
\end{document}